\documentclass[useAMS]{mn2e}
\usepackage{graphicx}
\def\hi{H\,{\sc i} }

\def\kmss{km~s$^{-1}$ }
\def\kms{km~s$^{-1}$}

\def\smppc2{${\rm M}_{\sun} {\rm pc}^{-2}$}

\def\kprime{K^{\prime}}

\def\cms{\hbox{cm s$^{-2}$}}
\title[Distribution of dark and luminous matter inferred from rotation curves]
{The distribution of dark and luminous matter inferred
from extended rotation curves.}
\author[R. Bottema and J.L.G. Pesta\~na]
{Roelof Bottema$^{1}$ and Jos\'e Luis G. Pesta\~na$^{2}$\\
$^{1}$Kapteyn Astronomical Institute, PO Box 800, NL-9700 AV Groningen,
The Netherlands, robot@astro.rug.nl\\
$^{2}$Departamento de F\'{\i}sica, Universidad de Ja\'en, Campus Las Lagunillas,
23071 Ja\'en, Spain, jlg@ujaen.es}
\begin{document}

\date{Accepted: date1. Received: in original form:}
\pagerange{\pageref{firstpage} -- \pageref{lastpage}} \pubyear{2015}

\maketitle

\label{firstpage}
\begin{abstract}
A better understanding of the formation of mass structures in the universe
can be obtained by determining the amount and 
distribution of dark and luminous matter in
spiral galaxies. To investigate such matters
a sample of 12 galaxies, most with accurate distances, has been
composed of which the luminosities are distributed regularly over a
range spanning $2\frac{1}{2}$ orders of magnitude.
Of the observed high quality and extended rotation curves
of these galaxies decompositions have been made, for four different
schemes, each with two free parameters.
For a ``maximum disc fit'' the rotation curves can be
well matched, yet a large range of mass-to-light
ratios for the individual galaxies is
required. For the alternative gravitational theory of MOND the 
rotation curves can be explained if the 
fundamental parameter associated with MOND is allowed as a free
parameter. Fixing that parameter leads to a disagreement
between the predicted and observed
rotation curves for a few galaxies.
When cosmologically motivated NFW dark matter halos are assumed,
the rotation curves for the least massive galaxies can, by no
means, be reproduced; cores are definitively preferred over cusps. 
Finally, decompositions have been made for a pseudo isothermal 
halo combined with a universal M/L ratio.
For the latter, the light of each galactic disc and bulge has been corrected
for extinction and has been scaled by the effect of stellar population.
This scheme can successfully
explain the observed rotations and leads to sub maximum
disc mass contributions.
Properties of the resulting
dark matter halos are described and a ratio between dark and baryonic
mass of $\sim$ 9 for the least, and of $\sim$ 5, for the most
luminous galaxies has been determined, at the outermost measured
rotation.
\end{abstract}
\section{Introduction}
It is now a firmly established result that there is a large discrepancy 
between the visible, luminous mass in a galaxy and the dynamical mass.
Observed rotation curves, especially those derived from the neutral 
hydrogen line, remain mainly flat in the outer regions.
However, the light distribution, which generally decreases
exponentially predicts that these rotation curves should be declining. 
This discrepancy can be explained by invoking dark matter (DM) surrounding
the visible component as an extended more or less spherical mass component
(Sancisi \& van Albada 1987; Knapp \& Kormendy 1987; Trimble 1987).
The first detailed observations of extended \hi rotation 
curves (Bosma 1978; Begeman 1987, 1989) 
were made of intermediate sized galaxies. These systems generally
show a rotation which remains flat over a very large range of radii. 
Such a situation can only be generated if the rotational contribution
of the inner luminous component and that of the outer dark component
conspire to make a combined curve at a constant level (the ``conspiracy'', 
van Albada \& Sancisi 1986). Later on it became evident that less luminous
galaxies have rotations which continue to rise in the outer parts while
the most luminous, and certainly the galaxies with a more concentrated
light distribution, have a rotation which rises quickly near the centre
and declines slightly in the outer regions. In that way the conspiracy
does not exist any more (Casertano \& van Gorkom 1991; Persic et al. 1996). 

Naively one could reason that by subtracting the rotational contribution
of the luminous component from the total observed rotation, the rotational
signature of the dark halo remains. The principle is right, but in practice
although the radial functionality of the luminous rotation can be derived
quite accurately, the absolute contribution is difficult to quantify. 
The unknown luminous scale factor might be determined by making a least
squares fit of the total combined rotation to the observations. In most cases
such a fit converges to a maximal contribution for the luminous mass,
but a close inspection of the fit procedure shows that the
solution is highly degenerate: nearly equally good fits can be achieved
when exchanging dark for luminous matter (van Albada \& Sancisi 1986;
Dutton et al. 2005).
Consequently, when doing such fits, constraints have to be set to
either kind of matter. One of such constraints is the maximum disc
hypothesis which states that the amount of luminous matter should
be maximized. There are no principle reasons why this constraint should 
apply, but it has the useful feature that the procedure minimizes the
amount of dark matter which is needed. Thus making the dark matter problem
as small as possible. Furthermore is sets a firm upper limit to any
determined mass-to-light (M/L) ratio of a galactic disc or bulge. 

In principle the amount of mass in a disc (or bulge) can be determined
by measuring the velocity dispersion of the stellar content. 
Such observations of stellar discs and the subsequent analyses
are not straightforward and consequently reliable measurement have
so far only been obtained for a few samples of galaxies. 
Yet such studies now generate a mounting evidence that a disc 
is sub maximal. {From} measurements of stellar ve\-lo\-city dispersions of 
a sample of inclined galaxies Bottema (1993) determined that,
on average, the maximum contribution of a stellar disc
to the total rotation is 63\% with an error of some 10\%. 
These observations and findings have been confirmed by Kregel et al. (2005)
for edge-on disc galaxies, by Herrmann \& Ciardullo (2009) by
measuring the kinematics of planetary nebulae, and by
Martinsson et al. (2013) for more face-on systems. The latter 
analysis even seems to indicate that the disc rotational contribution
is smaller than the 63\% mentioned above. 
There is additional evidence for such a sub maximum disc from
a statistical analysis of rotation curve shapes in relation to the 
compactness of discs (Courteau \& Rix 1999).

An alternative assessment of the matter of disc contribution can be made
by considering mass-to-light ratios. The Initial Mass Function (IMF) seems
to be universal for a range of normal galactic conditions (Kroupa 2001). 
That implies that identical stellar populations have identical mass-to-light 
ratios. By considering a broad spectrum of galaxy formation scenarios and
a number of different population synthesis codes, Bell \& de Jong (2001)
establish a tight relation between the optical colour of a galaxy and its 
M/L ratio in every optical and near infra-red passband. This relation is
well established, but the absolute value of the M/L 
ratio depends on the adopted IMF.
If the low mass end of the IMF changes then also the M/L changes, typically
for a standard Salpeter IMF (Salpeter 1955) with slope 1.35, M/L $\propto
m_l^{-0.35}$ where $m_l$ is the low mass cutoff.
Thus, one is back at square one:
it is not possible to know the luminous mass contribution a  priori.
Anyhow, the relative functionality of Bell \& de Jong
is very useful because it allows a mass scaling of
the observed population, like the 
preliminary procedure used by Bottema (1997).
For instance,
a red population is relatively old and deficient in light and so its associated
mass should be scaled up. On the other hand, a blue population is young
and bright and its light has to be scaled down to get the appropriate mass.

Historically, when dealing with the distribution of dark matter
in the context of extended rotation curves, a radial density functionality
is used equal to that of a pseudo isothermal sphere 
(Carignan \& Freeman 1988, Begeman 1989). Objects of this sort are indeed
produced by gravitational instabilities of a universal fluid having 
density fluctuations of a specified plausible form (Silk 1987).
But recent detailed calculations and simulations of the formation of
mass structures in the universe seem to require dark matter of a
collisionless and cold kind. Such CDM (Cold Dark Matter) is able to explain,
with considerable success, the observed filamentary and honey-comb structures
of clusters and superclusters (Davis et al. 1985; Bond et al. 1996), appearing 
with the right sizes and at the right times. Collisionless matter entities get
shaped by continued interactions, merging, and violent relaxation
of smaller aggregates into larger structures. Inevitable that forms dark
halos with density profiles close to that of 
an NFW (Navarro, Frenk, \& White 1997)
shape. Such a profile has an inner functionality proportional 
to $R^{-1}$ (cusp) while
a pseudo isothermal halo has a central core with constant density. 
In any way, it seems natural to assume that dark halos of present day galaxies
should have a shape similar to that of an NFW profile.

Despite its success in explaining the large scale structures, CDM
predicts a number of details which are not in agreement with observations.
The three disagreements which stand out are: the expected cusps while cores 
are observed, the expected large number of dwarf satellites which should
orbit a larger galaxy, which are not there, and the expected triaxiality
of DM halos which is not observed. The first item of this list has
been considered and discussed extensively in the recent literature.
For example, there are a number of
observations of the rotation in small or LSB galaxies, where the DM should
dominate, which clearly show that the dark halos 
of these galaxies have cores (de Blok et al. 2001a,b, 2003;
Salucci et al. 2003; Simon et al. 2003; Blais-Ouellette et al. 2004;
Gentile et al. 2004). 
Several mechanisms have been invented to alleviate or modify
the CDM predictions (see Pe\~narrubia et al. 2012, and references
there in). One of the most promising seems to be cusp flattening by
supernova explosions after an idea by Read \& Gilmore (2005) and
studied extensively by Governato et al. (2010) and by Pontzen \&
Governato (2012). Yet, in order to accomplish this, baryonic matter has
to flow into the central regions of primeval galaxies until over there the
density becomes comparable to that of the dark matter. Star formation
has then to be suppressed until such densities are reached and 
subsequently an order of magnitude of the baryonic mass has to
be expelled by SN explosions, taking along the central DM to
the outer regions. It remains a matter of debate if this mechanism
can actually work or whether it is in agreement with observed
star formation histories and metallicity content of galaxies. 
For the smaller systems and certainly for dwarf spheroidals
the energy requirements contradict such a scenario (Pe\~narrubia
et al. 2012; Garrison-Kimmel et al. 2013).

As an alternative to dark matter the suggestion has been put forward
that the usual law of Newtonian gravity breaks down on the scale
of galaxies. In particular there is the proposal by Milgrom (1983) that
the effective law of attraction becomes more like $1/r$ in the limit
of low accelerations. This proposal designated as MOND: Modified Newtonian
Dynamics, has been successful in explaining certain aspects of the
difference between luminous and dynamical rotation on the scale of
galaxies and groups of galaxies (Sanders 1990; Sanders \& McGaugh 2002).
MOND has been generalized and moulded into a relativistic
theory with appropriate yet complicated field equations
(Bekenstein 2004). If this theory proves to be comparable
or superior to the DM paradigm it cannot be discarded as a toy model,
and it could be the expression of a more fundamental principle.

Let's go back to the real world. MOND fitting has been applied to
the rotation curves of galaxies in a number of studies. This fitting can
and has been done in a few different ways depending on the number of allowed
free parameters. In principle MOND is governed by one single acceleration
parameter $a_0$, which has to be equal for all structures in the universe. 
For galaxies this leaves as a free parameter the amount of Newtonian 
contribution to the rotation, which can be parameterized by the M/L ratio
of the luminous component. The contribution of the gasmass is fixed. 
Abandoning MOND as a fundamental theory, the acceleration parameter $a_0$ can
be considered as a free parameter too. This kind of fitting will be
referred to as MOND-like and in first instance it can be used to check
if indeed an equal value for $a_0$ exists. In practice and certainly before
accurately measured Cepheid distances became available, distance can be 
used as a third free fitting parameter, to within certain limits, of course. 
So in its broadest sense MOND fitting has three free parameters and
as such can explain every available rotation curve with ease. 

A first detailed analysis of the MOND method has been done by
Begeman, Broeils \& Sanders (1991, hereafter BBS) .
They selected a sample of only 10 galaxies with 
high quality and extended rotation curves. 
Most of these curves could be fitted with the MOND prescription using
a universal value of $a_0$ equal to 1.21 10$^{-8}$ \cms\ with only slight
adjustments to the Hubble distances. With one exception, being NGC 2841
which needed a MOND preferred distance twice as large as its Hubble distance. 
Well, one exception should confirm the rule. However, during the course
of the Hubble key distance project a few Cepheid distances became available
for galaxies in the sample of BBS. These distances are accurate to approximately
10\% and appeared to differ in some cases considerably from the Hubble 
distances used before. Notably NGC 3198 was further away and NGC 2841 appeared
to be closer than the preferred MOND distance. As analysed and discussed by 
Bottema et al. (2002, hereafter BPRS), it is really difficult to reconcile
the distances of both these galaxies with a universal $a_0$ parameter, unless
one or more Cepheid distances are considerably wrong.

A few more studies have been done investigating the applicability 
of MOND in general and its consistency with observed rotation curves
specifically. Different studies use different galaxy samples and
a different number of free fitting parameters and are only
generally comparable. 
For example de Blok \& McGaugh (1998) consider a sample
of 15 Low Surface Brightness (LSB) galaxies, which should reside
largely in the weak acceleration or MOND regime and conclude that
the MOND fit is excellent in 75\% of the cases. Swaters et al. (2010)
reach essentially the same conclusion for a larger sample of LSB
galaxies. Gentile et al. (2011) consider 12 nearby galaxies for
which rotation curves are available from the THINGS sample. For a
MOND-like fit these authors find an average value for $a_0$ 
essentially equal to the standard number of BBS. Fixing that value
three quarters of their sample can well be fitted; leaving 
distance as an additional free parameter all galaxies can
be accommodated, as noted above. In a recent study of
Randriamampandry \& Carignan (2014) for a sample of 15 nearby
galaxies, partly coinciding with the sample of Gentile et al,
while keeping the (M/L)$_{3.6\mu}$ ratio and the
value of $a_0$ fixed, find that only 60\% of the galaxies
can be made in agreement with the MOND prescription.

As part of a broader investigation of the distribution of dark matter
in the present paper, also MOND has been considered. 
In first instance distances are now assumed to be known accurately.
Then MOND-like fits are made to investigate how
MOND with two free parameters compares with the procedure of 
adding DM with the same degree of freedom. 
In second instance it has been checked if the MOND philosophy works:
can all rotation curves be explained with one 
universal $a_0$ value for reasonable
adjustments of the distances. 
 
Presently the approach of BBS has been taken over: quality is
preferred over quantity. A sample of 12 galaxies has been compiled of which
7 are in common with the sample of BBS. 
The requirement of a rotation curve extending
beyond the edge of the optical disc is essential. If not, then certainly
for the more massive galaxies, any fit to, or determination of the
DM properties is usually possible, but meaningless. Four different schemes
have been employed in decomposing the observed rotation curves. Each scheme
uses a fitting method with two free parameters and are in that sense
comparable. An extensive analysis is made of the quality and probability
of each scheme; some work better, others are even ruled out. 
This all in an effort to find the most likely distribution of dark matter
and to restrict the range of possible distributions. 

This paper is organized as follows. In section 2 the sample is discussed
and peculiarities of each galaxy have been specified. 
Section 3 gives a description
of the extinction and population corrections in order to get a reliable
amount of representative light for each system. In section 4 the rotation
curve decomposition procedure is described in a general way.
That procedure has been applied to the sample
in section 5 for a maximum disc fit with pseudo isothermal halo.
In section 6 MOND-like fitting and subsequent analysis of the MOND philosophy
is presented. In order to relate the cosmologically motivated NFW halos to
present day galaxies it is necessary to consider the process of 
adiabatic contraction. That process is described in section 7 and then
incorporated into the fitting of NFW halos presented in section 8. 
In case of the more physically and observationally motivated scheme
of an equal universal M/L ratio the fitting results are given in section 9,
where also an extensive discussion is presented concerning the uncertainties
and the Tully-Fisher relation (Tully \& Fisher 1977). 
Finally in section 10 a more
general discussion and point by point conclusions have been put together. 
Throughout, a Hubble constant of 75 km s$^{-1}$ Mpc$^{-1}$ has been adopted. 

\section{The sample of galaxies}
A sample of 12 galaxies has been composed based on the following
criteria:
\smallskip

\begin{enumerate}
\item
Rotation curves derived from a two dimensional radial velocity field.
\item
The velocity field should be regular and non-distorted, without large
scale asymmetries.
\item
It should extend beyond the optical edge ($\sim$ 5 scalelengths) of
the disc.
\item
Inclinations larger than 50\degr, in order to derive, independently,
the inclinations and rotation velocities from the velocity field.
\item
Inclinations less then 80\degr, to avoid composite velocities along
the line of sight.
\item
Photometry is available in a band redder or equal to R.
\end{enumerate}
\smallskip
and based on the following preferences:
\smallskip
\begin{enumerate}
\item
Rotation curves extending as far as possible beyond the optical edge.
\item
A well determined distance, five galaxies have measured Cepheid distances.
\item
A velocity field and related rotation curve which are sampled by a 
sufficiently large number of independent data points.
\item
Optical emission line radial velocities in the inner regions when
at those positions beam smearing of the \hi data might be present.
\item
Measured stellar velocity dispersions of the disc to have
a handle on the local mass density.
\item
To generate a sample which is evenly spread over a large mass range.
\end{enumerate}
Without any doubt there will be more than the
12 galaxies in the sample which satisfy the requirements,
but we did not have the intention to be complete. 
The criteria and preferences above, can only be met when
galaxies are observed in the neutral hydrogen line with an
interferometer. A rotation curve is usually derived by a least
squares fit to the velocity field of a collection of rings with 
radially variable inclinations, lines of nodes,
and circular velocities. This procedure is described
in detail by Begeman (1987, 1989).
When \hi data have been replaced by optical emission 
lines in the inner regions, the latter have on occasion
been radially resampled to match the \hi sampling. 
That ensures an equal radial weighting when doing the
RC fit, which is a choice. Having more data points at a 
specific region may slightly change the relative 
contributions of the disc, bulge, and
dark halo, but will not change any of the main 
results (see Blais-Ouellette et al. 1999).

The parameters and properties of the sample have been
collected in the period 2003 to 2007 after which the analysis
as presented in the remainder of this paper has been carried out. 
Because of personal reasons the work on this paper 
had to be abandoned for a while.
During that time results of the THINGS (The \hi Nearby
Galaxy Survey; de Blok et al, 2008) had become available.
Of the 12 galaxies in the present sample 6 are in common
with the THINGS sample: NGCs 2403, 2841, 2903, 3198, 7331 and
DDO 154. When comparing the \hi rotation curves one can conclude
that THINGS has a better radial sampling because observations
have been done at a higher spatial resolution.
But in general the observations do not have the signal-to-noise
of the rotation curves already considered. The THINGS data then
have a shorter radial extent for all the galaxies in common. 
Especially, in order to determine
the DM distribution, these outer data points of the RC are of utmost
importance and therefore, using the THINGS rotation curves 
instead of the present ones would not improve on the results.

The rotation curves appeared to be nearly identical at
the positions where both are determined,
except for DDO 154 where for
the inner 3 kpc THINGS gives a larger rotation by a few \kmss
compared with the rotation curve of Carignan \& Purton (1998) which had
been used. Likely the latter suffers from beam smearing caused by
the very limited spatial resolution. Therefore it appeared useful
to take over the THINGS rotation curve for this specific case.

Over the past few years 3.6$\mu$ infrared photometry has become
available for a number of galaxies from SINGS (Kennicutt et al. 2003),
which has been used by de Blok et al. (2008) to characterize
the luminous radial density distribution. When comparing these
radial luminosity profiles with the ones of Kent (1986, 1987, hereafter K86 and K87) 
and Wevers et al (1986),
in the Red bands one is struck by the similarity. This should not
be obvious; radial absorption, population, and metallicity gradients
might generate a difference between the red and infrared profiles. 
An explanation is left to others. Anyway, using the 3.6$\mu$ profiles
instead of the present red profiles would not make a difference
for the subsequent analysis or results. On the other hand, using
the infrared luminosities would pose a serious problem for
a population and a necessary metallicity correction
for galaxies as a whole. Stellar population analyses in the infrared
are exceptionally difficult and moreover the emission may
be contaminated by a contribution of PAHs (polycyclic aromatic
hydrocarbons). For photometry in the Red, population corrections are
well described and the light is barely dependent on metallicity.

For all the galaxies a short description is now given
of the adopted distances, photometry, and rotation curves. 
Uncertainties, peculiarities and references are quoted. 
In Table~1 a number of important parameter values are summarized.

\subsection{NGC 2841}
A distance of 14.1 $\pm$ 1.5 Mpc has been determined by HST
measurements of Cepheids (Macri et al. 2001).
Photometry is available by K87 in the Thuan \& Gunn (1976)
r-band giving both, the photometric profile and total magnitude. 
Because there is a substantial difference in observed ellipticity
between the bulge and disc a decomposition of the light of these
components according to Kent's procedure is well determined.
The \hi rotation curve has initially been 
measured by Bosma (1981) and is later refined
by Begeman (1987). The gas distribution is symmetric, but the
kinematics displays a sizable amount of warping, making the determined
rotation in the outer regions slightly less certain. The rotation
curve remains more or less at a constant level all the way inwards
to the centre. The THINGS rotation curve reaches out to 36 kpc and
displays a sizable uncertainty between 25 and 36 kpc,
while the present one goes out to 64 kpc, both are fully consistent. 
 
\subsection{NGC 3992}
NGC 3992 is one of the most prominent members of the Ursa Major
cluster of galaxies. Tully \& Pierce (2000) determined a distance
of 18.6 Mpc to this cluster, which will be adopted as the distance
for NGC 3992.
There is photometry in the BRI$\kprime$ bands by Tully et al. (1996)
of which the I profile is used to calculate the disc rotation and
the R-band for the total light. The signature of the bulge is clearly present
in the observed radial light profile and consequently the bulge to disc
decomposition is straightforward.  
The \hi rotation curve has been measured
in detail by Bottema \& Verheijen (2002), who also give a rotation curve
decomposition. This galaxy has a bar and this bar region is devoid of
gas so that no rotation is available for the inner regions. 

\subsection{NGC 7331}
The HST Cepheid distance to this galaxy is 14.72 $\pm$ 0.60 Mpc
(Hughes et al. 1998; Freedman et al. 2001). Photometry is given by K87
in the r-band, but unfortunately the ellipticity of the central bulge
and outer disc appear to be equal and Kent's decomposition procedure
cannot be used. A detailed stellar and emission line kinematical study
of the galaxy has been made by Bottema (1999). In that paper
a decomposition into separate bulge and disc is presented, partly based
on  the observed stellar absorption line profiles. The bulge appears
to be quite dominant and nearly spherical while the disc has an 
inclination consistent with the \hi kinematical value of 75\degr. 
Unfortunately an error has turned up in Table~3 of Bottema (1999).
For the favourite decomposition labelled ``lpd'' the total amount
of disc and bulge light have been interchanged. The total light of disc
and bulge should be 11.4 $10^9 L^I_{\sun}$ and 12.7 $10^9 L^I_{\sun}$
respectively and the b/d ratio is 1.1 instead of 0.9. The radial profile
in the I band is given by Prada et al. (1996) which is used presently,
although we have some doubt regarding the absolute calibration of 
this I-band photometry. Total R-band light is derived from K87. 
The \hi rotation is determined by Begeman (1987), extending out
to a radius of 37 kpc, and he supplements the 
rotation in the inner region with data of Rubin et al. (1965). However,
as demonstrated by Bottema (1999) the gas kinematics in the inner regions
of this galaxy is very unusual and certainly not representative for
the gravitational potential. Therefore, at those positions the rotation
as inferred from the stellar kinematics is adopted. The THINGS rotation
curve only extends from a radius of 4 kpc to 24 kpc. The measured stellar 
velocity dispersion of the disc suggests that the contribution of the disc to
the total rotation is small. This measurement is somewhat uncertain, however,
because of the presence of a considerable amount of bulge light. 

\subsection{NGC 2998}
NGC 2998 is the most distant galaxy of the sample. The radial velocity
corrected for Virgo-centric flow gives a Hubble distance of 67.4 Mpc. 
Considering the deviations from the Hubble flow, this distance is
nearly as accurate as the Cepheid distances for the galaxies more close by. 
Photometry is presented in K86; the galaxy is close to exponential
with only a minor bulge. For Kent's bulge-disc decomposition method
the bulge has a luminosity of only 2\% of that of the
entire galaxy.
Therefore, presently, the whole galaxy is considered as a disc structure,
which effectively means that the mass-to-light ratio of the bulge
is equal to that of the disc. 
An accurate \hi rotation curve has been determined by Broeils (1992a),
which has unfortunately not been published in the refereed literature. 
The velocity field is very symmetric and regular and the gas extends
far beyond the optical edge. Because of its large distance the \hi
structure is not as well resolved as that of the more nearby galaxies
and consequently the rotation curve is more sparsely sampled. 
Moreover, the observed kinematics in the inner regions is affected by
beam smearing and cannot be used. 

Absorption line spectroscopy is obtained along the whole major
axis of NGC 2998, with a (1$\sigma$) velocity resolution of 
25 \kms\ (Bottema \& Kregel 2014). From that, stellar radial
velocities and stellar velocity dispersions have been derived
extending to approximately two and a halve optical scalelengths. 
In addition the emission lines of H$\beta$ and [O~III] 5007 \AA\ are
observed, which have been combined to generate the rotation
in the inner 40\arcsec, there where the \hi data are not useful.
The emission line kinematics is symmetric, but slightly irregular,
in the sense of showing some corrugation by less than 10 \kms. 
The stellar velocity dispersion decreases radially as expected
for an exponential disc. Depending on the exact parameterization of
the disc, a maximal contribution of the rotation of the stellar
component to that of the total rotation is calculated. This value
ranges between 0.64 and 0.72 with an error of 10\%.

\subsection{NGC 2903}
The Hubble distance to this galaxy is 6.3 Mpc, while a distance
estimate based on the brightest stars (Drozdovsky \& Karachentsev 2000) is
8.9 $\pm$ 1.9 Mpc. 
Presently we take as distance the average of the two: 7.6 Mpc. 
Photometry is measured by K87 giving the profile and total light. 
The rotation curve is composed of H$\alpha$ emission line data
by Marcelin et al. (1983) for the inner 100\arcsec\ and \hi data by
Begeman (1987) beyond, out to 29 kpc. The H$\alpha$ radial velocity measurements
have been transformed into a rotation curve using the same inclination of
62\degr as for the \hi kinematics. The THINGS rotation curve goes out to 25 kpc. 

\subsection{NGC 3198}
The distance of 13.8 $\pm$ 0.5 Mpc is derived from the HST 
Cepheid observations of this
galaxy (Kelson et al. 1999; Freedman et al. 2001), and is considerably
larger than the Hubble distance of 9.4 Mpc. Again, photometry by
K87 is used of which the validity to represent the mass distribution
is confirmed by $\kprime$ band photometry by BPRS. 
Because the \hi rotation curve (Begeman 1989) reaches very far out NGC 3198 has become
the classic case of a spiral galaxy evidencing a large mass
discrepancy in its outer regions (van Albada et al. 1985).
Numerous re-observations of the \hi and re-determinations of the 
rotation curve have been made. At the flat part none of these
warrant a change to what had already been determined by Begeman (1989).
Additionally all observations end at the radius of nearly 45 kpc
probably because at larger galactocentric distances the \hi gas becomes ionized. 
The rotation curve of THINGS ends at a radius of 38 kpc. However
that curve seems to show a somewhat lower rotation in the 
inner rising part, for R $\la$ 5 kpc. Whether this is real or might be
caused by a bar feature, needs to be confirmed. In this inner region
of NGC 3198 matters appear to be quite complicated and a thorough
analysis could easily fill a complete paper and is certainly beyond the
scope of this study. Presently Begeman's curve has been taken over
entirely with its relatively large errors in the inner region, nicely representing
the current uncertainty.
 
For its size and mass NGC 3198 has indeed a relatively large
amount of gas present (see Table~1).
Stellar velocity dispersions of the disc of this galaxy have been
measured by Bottema (1988), which for an assumed average disc
thickness imply a sub maximum disc contribution to the rotation curve. 

\subsection{NGC 2403}
This galaxy has a distance based on Cepheid variables
(Freedman \& Madore 1988; Freedman et al. 2001). Photometry by
K87 and \hi rotation curve by Begeman (1987). Like for NGC 3198 this
galaxy has also been re-observed in the \hi on numerous occasions,
which has never led to any alteration of the rotation curve used here. 

\subsection{NGC 6503}
A distance of 5.2 $\pm$ 1.1 Mpc is 
assumed based on measurements of the 
luminosities of the brightest blue stars
(Karachentsev \& Sharina 1997), being
in agreement with the Hubble distance of 4 Mpc and Tully-Fisher
distance of 6 Mpc (Rubin et al. 1985). The luminosity profile
is a composite of photographic Kodak IIIa-F measurements of 
Wevers et al. (1986) in the outer regions and R-band CCD photometry
by Bottema (1989) in the inner regions. All is converted to
R-band magnitudes. The \hi rotation curve is from Begeman (1987).
As for NGC 3198, stellar velocity dispersion observations (Bottema 1989)
imply that the mass contribution of the disc is sub maximal. 

\subsection{NGC 5585}
In this case the Hubble distance of 6.2 Mpc is the most reliable
value available. Photometry in the B,V, and R-bands has been
obtained by C\^ot\'e et al. (1991) of which the R-band profile is used
to calculate the rotation curve of the disc. This profile has
been integrated to give the total R-band luminosity.
An \hi rotation curve is determined by C\^ot\'e et al. from a regular
though slightly warped velocity field. For radii within 120\arcsec\
the rotation of Blais-Ouellette et al. (1999) based on H$\alpha$ Fabry-Perot
observations is used. That inner rotation curve shows a specific bumpy 
feature which can be matched with the moderate central cusp in the photometry.

\subsection{NGC 1560}
The distance to this galaxy is taken as the average of three values.
At first the Hubble distance of 3.25 Mpc. Secondly a distance of
2.5 $\pm$ 0.1 Mpc (Lee \& Madore 1993) based on bright stars and
Tully-Fisher relation, and thirdly Krismer et al. (1995) give a
distance of 3.5 $\pm$ 0.7 Mpc based on the TF relation. The average
of the three amounts to 3.1 Mpc.  
Broeils (1992b) presents 
photometry in the B-band and Swaters \& Balcells (2002)
in the R-band. The profile of the latter authors declines 
slightly steeper and was used to represent the mass distribution
of the disc. Total R-band luminosity
also from Swaters. For radii beyond 120\arcsec\ the \hi rotation
curve of Broeils (1992b) is adopted. Because this galaxy is rather
edge-on ($i \sim$ 80 to 82\degr) the available \hi may be compromised
by beam smearing and integration effects. Therefore long slit H$\alpha$
data of de Blok \& Bosma (2002, which have kindly been made available by
these authors) have been converted to rotational velocities.
For such a high inclination galaxy slit data are not ideal. 
Yet this galaxy is large on the sky and one H$\alpha$ RC data point
is the average of a lot of individual observations, showing quite
some scatter, as expected. This average and associated error
are judged to be sufficiently accurate.
Between 120 and 200\arcsec\ the H$\alpha$ and \hi rotations are equal, for
radii less than 120\arcsec, the emission lines indicate a slightly larger
rotation by approximately a few to 10 \kms. Therefore, at those
positions the H$\alpha$ is used for the rotation.  

\subsection{NGC 3109}
For this galaxy a Cepheid distance is available of 1.36 $\pm$ 0.10 Mpc
by Musella et al. (1997). These authors use the same distance to
the LMC as the HST Cepheid distance key project and is in that sense
comparable. Yet there may be small systematic differences between
the two methods. Recently, Soszy\'nski et al. (2006) determined a 
distance of 1.30 $\pm$ 0.04 Mpc by extending the Cepheid observations to
the infra-red. This confirms the value of Musella et al. and the
difference is so small that we maintained the value of 1.36 Mpc.
Uncalibrated I-band photometry has been obtained
by Jobin \& Carignan (1990) of which the radial profile was used to
calculate the rotation of the disc. A total R-band luminosity was
derived from the ESO LV catalogue (Lauberts \& Valentijn 1989) by
extrapolating the R-band aperture value at R$_{26}$ equal to the B-band
functionality. 

The construction of the rotation curve is somewhat comp\-li\-cated. 
{From} the I-band photometry, assuming $q_0 = 0.11$ one has an 
inclination of 75\degr, which seems well determined.
\hi observations of Jobin \& Carignan suggest an inclination of
70\degr\ but is rather uncertain. Therefore the errors on the 
\hi rotation have been increased to include this uncertainty
in the inclination. For the inner regions ($<$ 350\arcsec) the rotational data
have been supplemented with H$\alpha$ Fabry-Perot observations
by Blais-Ouellette et al. (2001). In that paper an inclination
is used of 88\degr, which is clearly in contradiction with other
determinations. Therefore the rotation of Blais-Ouellette et al. has
been converted to an inclination 75\degr. Moreover, the quoted
errors appeared to be unrealistically small and have for all H$\alpha$ rotational
data points been increased to 5 \kms. That should then also include
any uncertainties associated with the asymmetric drift correction. 
This might all seem a bit tricky, but for a galaxy with such a large
inclination small changes in the inclination never substantially
affect the rotation.  

\subsection{DDO 154}
A description of distances, photometry, and rotation can
be found in the paper of Carignan \& Beaulieu (1989). 
The distance of 3.8 Mpc is based on a combination of brightest
star considerations and association with the Canes Venaticorum I cloud. 
Photometry is available in the B,V, and R bands. An additional 
measurement in the R-band by Swaters \& Balcells (2002) is consistent, both
concerning the profile and total luminosity. 

The \hi rotation curve of Carignan \& Beaulieu exhibits a decline
beyond a radius of $\sim$ 300\arcsec. In a following study by
Carignan \& Purton (1998, hereafter CP98) additional observations reveal a more
extended low level \hi structure. The outer \hi distribution is lopsided
but the velocity field appears rather regular. Again the derived
rotation curve starts to decline abruptly and continues so out
to a radius beyond 540\arcsec. The derivation of the 
rotation curve by these authors seems solid.
As mentioned above, the rotation curve of THINGS qualifies
to replace the one of CP98 because their
inner data points have probably been affected by beam smearing.
Moreover the errors which have been given are unrealistically small. 

Then we encountered a problem. In the paper of De Blok et al. (2008)
two different rotation curves for DDO 154 are presented, 
one slightly rising in the outer parts in their figures 15 and 46,
and one nearly constant or decreasing slightly in their figure 81. 
The different rotation curves have been derived for a differently
adopted inclination functionality between 300\arcsec\ and 400\arcsec,
both of which are consistent with the THINGS observations. 
For this galaxy at those positions the tilted rings used to
fit the velocity field are only partially
and irregularly filled with data on hence an independent inclination determination
is not possible. The observations of CP98 extend further out, 
until a radius of approximately 540\arcsec; their velocity field appears to
be regular over there with an inclination nearly constant at 60\degr,
which coincides with the choice of De Blok et al. in their figure 81. 
Consequently that rotation curve of THINGS is to be preferred
and as such is completely compatible with the rotation curve of 
CP98.  In this study that rotation curve has been
taken over, supplemented with the data points at 450\arcsec\ and
540\arcsec\ of CP98 with an error increased to a
realistic value of 5 \kmss for an adopted inclination uncertainty of 10\degr. 

The radial distribution of the gas has been taken from 
Carignan \& Beaulieu (1989) for the inner regions. It has been extended
to larger radii by a smooth matching to it with the distribution
given in Fig. 4 of CP98.

\begin{figure}
\resizebox{\hsize}{!}{\includegraphics{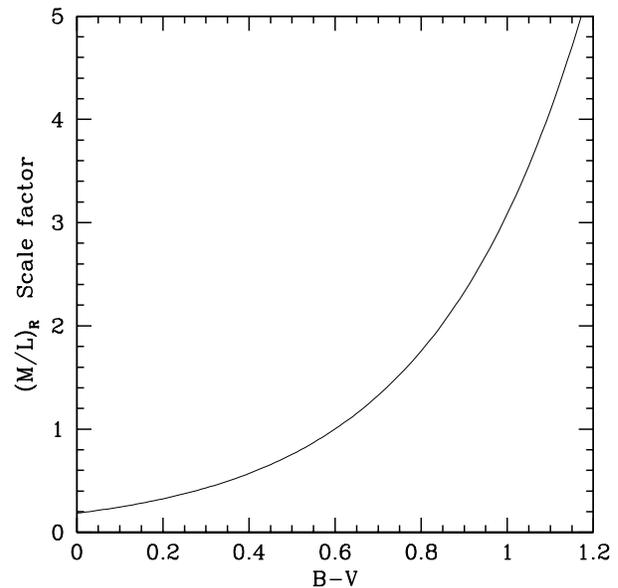}}
\caption{Mass-to-light ratio in the R-band as a function of
B-V colour (Bell \& de Jong 2001), lowered by 15.5\% to give a value of
1.0 at B-V~=~0.6. This relation is used to scale the amount
of light in a galaxy to a representative value for the
amount of mass.}
\end{figure}

\section{Total light in R, extinction and population corrections}
To obtain an amount of light which is representative for the amount of
luminous matter in a galaxy corrections have to be made to the
observed total light, in this case in the Kron-Cousins R-band. 
There are two main corrections to be made; a correction for
absorption to be called extinction correction, and a correction
for population. The latter is needed because a younger population
generates more light compared to an older population for the same
amount of stellar mass. Uncorrected parameters have been given
the subscript ``obs'', those corrected only for extinction ``ec''
and those corrected for extinction plus population are given
the subscript ``epc''.

\subsection{Galactic extinction}
Extinction from the local Galaxy, designated by $A^b$ is taken
from Schlegel et al. (1998). Its values for the sample of
galaxies can be found, with other forthcoming corrections, in
Table~2. 

\subsection{Internal extinction to face-on}
A correction for this extinction $(A^{i-0})$ is given by Tully et al.
(1998) and depends on the absolute luminosity of a galaxy corrected
for Galactic and internal extinction $(M_R^{b,i-0})$ and on the
observed aspect ratio $a/b$ as
\begin{equation}
A_R^{i-0} = {\gamma}_R \; {\rm log} \left( \frac{a}{b} \right),
\end{equation}
with
\begin{equation}
\begin{array}{rl}
{\gamma}_R  &=\; -0.24\; (16.06 + M_R^{b,i-0} )\\
            &=\; 0\;\; {\rm if}\;\; M_R^{b,i-0} > -16.06.
\end{array}
\end{equation}
This correction has been applied separately to the disc
and the bulge (see Table~2).

\subsection{Intrinsic extinction of a face-on galaxy}
This matter is uncertain because it has never been investigated
in a systematic way. Therefore we can only give a reasonable
estimate. Tully \& Fouqu\'e (1985) and \hbox{Verheijen} (2001) use an 
amount of intrinsic extinction $(A_R^{i=0})$ of 0.21 mag. for an 
average galaxy in their samples. Comparing these samples with 
the galaxies in this paper such an average galaxy has approximately
the luminosity of NGC 3198. It is further assumed that the 
intrinsic extinction is smaller for less luminous galaxies,
analogous to the extinction correction to face-on.
Adopting $A^{i=0} = $ zero for $M_R^{b,i-0} > -16.0$ we use
for $M_R^{b,i-0} < -16.0$:
\begin{equation}
A_R^{i=0} = -0.042\; M_R^{b,i-0} - 0.672,
\end{equation}
for discs and $A_R^{i=0} = 0$ for bulges. 

\subsection{Population correction}
Bell \& de Jong (2001) calculate mass-to-light ratios for galaxies
as a function of their colours. The existence of such a colour
versus M/L ratio relation can be understood because a younger, lower
M/L ratio population is relatively bluer compared to an older,
larger M/L ratio population. Bell \& de Jong demonstrate that such
a relation is largely independent of galaxy evolution scenarios and
on the employed population synthesis code. In the non near infra-red
passbands it barely depends on metallicity of the population. 
For the M/L ratio in the R-band and B-V colour Bell \& de~Jong give a relation of
\begin{equation}
{}^{10}{\rm log}(M/L_R) = -0.66 + 1.222 (B-V)
\end{equation}
As for all population synthesis analyses the M/L ratio is
only known up to a certain factor. This factor depends on the
assumed low mass end of the IMF and might be as large
as two. We can therefore not use Eq.~(4) to derive
M/L ratios, but we shall use it to correct the amount of R-band
light for the excess amount of light of a young population,
or deficiency of light of an old population. A fiducial B-V of
0.6 is chosen for which such a population correction is zero.
Then, using Eq.~(4) with M/L lowered by 15.5\%,
in Fig.~1 the scale factor is given
with which the amount of light has to be multiplied
to obtain the representative amount of mass.

In section 9 a successful fit to all rotation curves can
be achieved by using an equal mass-to-light ratio in R,
corrected for extinction and population, having a value of 1.0.
This means that the scale factor presented in Fig. 1 is then
exactly equal to the (M/L)$_R$ ratio as a function of B-V colour
one needs for galaxies in general. As a consequence the relation
as calculated by Bell \& de Jong, on the basis of hitting the
default lower mass cutoff in the population synthesis codes of 0.1 M$_{\sun}$
is then 15.5\% too large. If, instead, a lower mass cutoff at
0.15 M$_{\sun}$ would have been taken, their relation would be
spot on the relation given in Fig. 1, and thus on what
the successful rotation fit scheme of section 9 implies.

B-V colours of galaxies can be found in the RC3 
(de Vaucouleurs et al. 1991), but in order
to be usable in Eq.~(4) should be corrected for absorption. For that
we used the recipe given in the RC3, to obtain the parameter
designated as $(B-V)_T^0$. Since rotation curve decompositions will
be made with a separate disc and bulge, it appeared 
necessary to apply the extinction and population corrections
separately to these components too. Unfortunately the 
$(B-V)_T^0$ values are not listed separately, but can be retrieved.
When the total observed $(B-V)_{\rm obs}$ is given (RC3), the ratio
of disc to bulge light has been determined (Table~1), and the
observed $(B-V)_{\rm obs}$ of the bulge is known, the observed
$(B-V)_{\rm obs}$ of the disc can be derived. The B-V of the bulge
has been assumed to be equal to the B-V colours of the smallest
apertures given by Longo \& de Vaucouleurs (1983). Subsequently
the separate $(B-V)_{\rm obs}$ colours have been converted to
$(B-V)_T^0$ according to the RC3. The results of the population
correction procedure are given in Table~3 for the sample of galaxies.
Note that the three bulges are relatively red, the population is 
therefore old and light has to be scaled up considerably. 

For all the forthcoming fits two kind of mass-to-light ratios will
be presented; as observed, $(M/L)_{\rm obs}$ meaning with
no correction at all and $(M/L)_{\rm epc}$ with both, the
extinction and population correction applied. In Sect. 9 a
decomposition is presented using an equal $(M/L)_{\rm epc}$
for all luminous galaxy components. 

\begin{figure*}
\centering
\includegraphics[width=16cm]{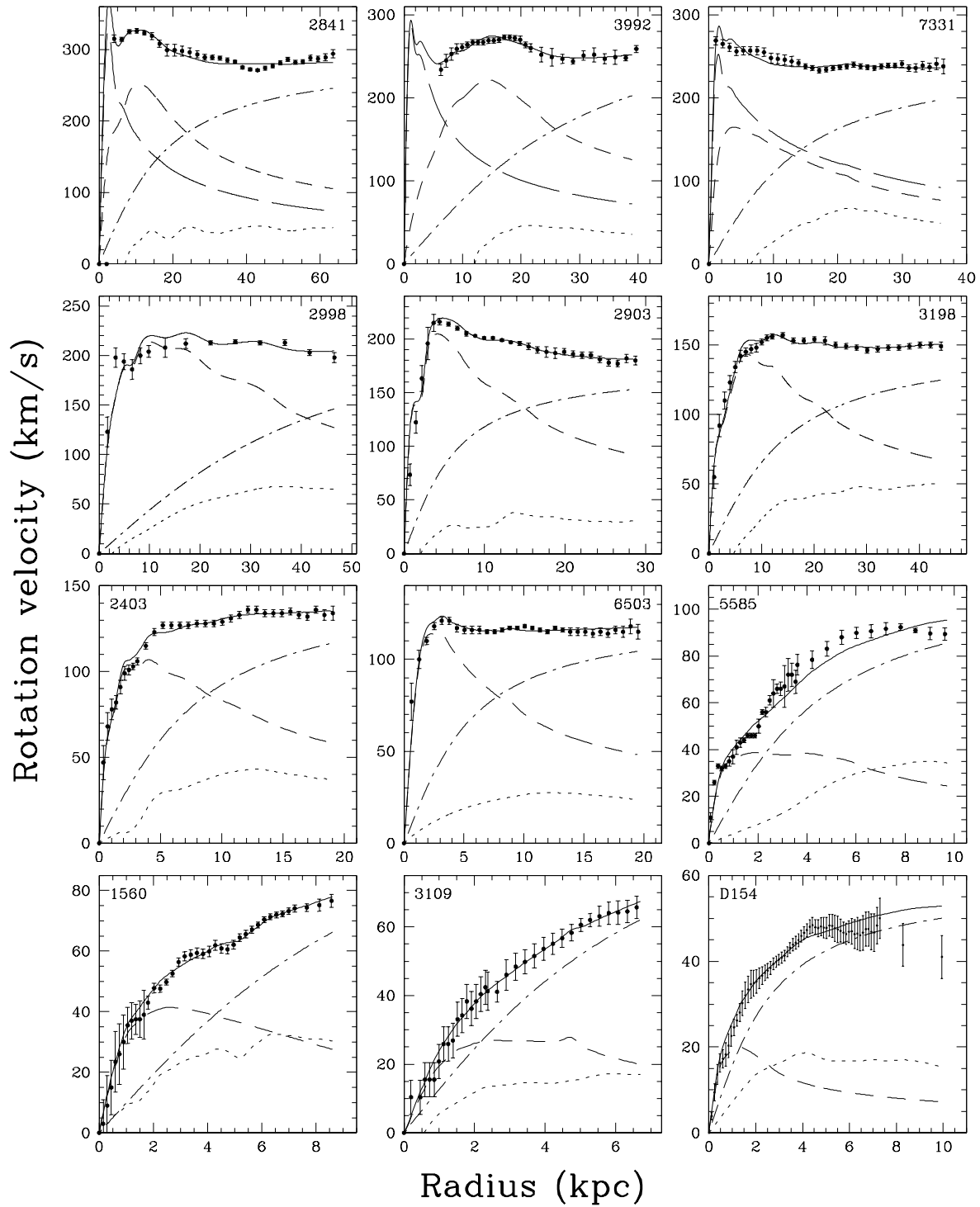}
\caption{Maximum disc fits to the observed 
rotation curves (dots) of the 12 galaxies
in the sample. Here, and in forthcoming similar plots (Figs 4, 7, and 9), the lines
are coded as follows: full drawn is the fit, dotted, short--dashed, long--dashed and
dash--dot lines represent the rotation of the gas, the
disc, the bulge, and the pseudo isothermal dark halo, respectively. The observations
can be matched quite accurately, except for NGC~5585 and DDO~154 which
show a small discrepancy.}
\end{figure*}

\section{Rotation curve fitting}
For every galaxy a model rotation curve is computed as the
squared sum of the individual contributions of the disc, bulge,
gas, and dark halo. The contribution of the gas is fixed by
observations of the neutral hydrogen gas, of which the amount
has been multiplied by a factor 1.4 to account for the presence 
of Helium and a small amount of ionized hydrogen. The disc and
bulge light distribution are measured by the photometry of the galaxy. 
Scaling by the M/L ratio then gives the mass contribution.
For the dark halo generally an analytical density functionality
is assumed described by one or two parameters. Thus the composite
model rotation curve has as free parameters the M/L ratio of the
luminous components and the descriptives of the dark halo. 
This composite curve is then fitted  in a least squares sense
to the observed rotation curve, a procedure often referred to
as decomposition of the rotation curve. 

In that way the best fit is designated as the situation of 
a minimum ${\chi}^2$ value. This procedure is scientific in
the sense that it is reproduceable and therefore is preferred
over, for example, an estimated fit by eye. Yet one has to be careful when
applying this method. A least squares fitting procedure
assumes that the fitting function is known a priori and that
the data points scatter in a Gaussian way around that function. 
For rotation curves that is not valid. Firstly a rotational
functionality for the halo is adopted which need not be correct. 
Secondly the procedure to determine the rotation is an approximation
in the sense that azimuthal symmetry is assumed with no in 
or outflow. For example spiral arms can produce small irregularities,
which may lead to small systematic deviations from the actual 
rotation law. In addition one has to be careful when "feeding"
the fitting procedure. At positions with a high degree of sampling
the fit is forced to a higher weight. Therefore, presently, rotation curves
have, on occasion, been resampled to a nearly uniform radial distribution
to give equal weight over the entire radial extent. 
Because of these matters the resulting minimum
${\chi}^2$ value is only a limited indicator of the quality of
the fit and it cannot be concluded from that value only
that a certain dark halo functionality is better or worse. 

The radial luminosity profiles of bulge and disc are adopted to
indicate the radial mass distribution. This assumes that the
M/L ratio is constant in the galaxy which can be justified
by the observed small radial colour gradients, in general
(de Jong 1996), at least for normal quiescent systems.
Nevertheless, in order to minimize dust and population effects and 
so to be close to the real mass distribution, a profile in a passband
as red as possible is preferred. Currently for all galaxies 
of the sample the profile is in a (R)ed or near
(I)nfrared passband (see Table~1).

For the discs the observed radial mass distribution is 
combined with an adopted sech$^2(z/z_0)$ vertical mass distribution
with $z_0$ being 0.2 times the value of the radial scalelength
(van der Kruit \& Searle 1981, 1982). The disc rotation curve is
then calculated following Casertano (1983). For the bulge which is
assumed to have a spherical distribution the rotational
velocity can be calculated using the equations on page 1310
of K86. Both for the disc and the bulge the rotation
curves can then be scaled up with the unknown M/L ratio. 
For the gas the radial mass distribution is observed directly. 
To calculate the rotation a thin vertical distribution is
assumed. Finally the dark halo. In this paper a few density
functionalities have been investigated usually parameterized
by two values. A detailed description will be given where
appropriate. The calculation of the rotations curves and the fitting
procedures have been performed using the routines {\sc rotmod}
and {\sc rotmas} in the {\sc Gipsy} package. 

In practice it appears that the fitting procedure is degenerate;
one generally cannot determine the M/L ratio and dark halo
parameters simultaneously. Consequently
it is necessary to impose certain constraints. In the remainder
of this paper, in principle, constraints are made such that
there are always two free parameters left (or three with a bulge);
in that sense the fitting schemes are then comparable. 

\section{Maximum disc and bulge fits}
In general the maximum disc
constraint is associated with a dark halo 
functionality of a pseudo isothermal
sphere (Carignan \& Freeman 1988). For consistency that functionality
will also be used presently. Its density distribution ${\rho}_h$
takes the form
\begin{equation}
{\rho}_h = {\rho}_0 \left[ 1 + \frac{R^2}{R^2_{\rm core}} 
\right]^{-1} ,
\end{equation}
with an associated rotation $v_{\rm p.iso}$ of
\begin{equation}
v_{\rm p.iso} = v^h_{\rm max} \sqrt{1 - \frac{R_{\rm core}}{R}
{\rm arctan} \left( \frac{R}{R_{\rm core}} \right) },
\end{equation}
where the halo rotation becomes asymptotically flat at
a maximum value of $v^h_{\rm max}$ which is related to the
central density ${\rho}_0$ and core radius $R_{\rm core}$ as
\begin{equation}
v^h_{\rm max} = \sqrt{ 4 \pi G {\rho}_0 R^2_{\rm core} }.
\end{equation}
Note that for small radii ${\rho}_h \approx$ constant and 
$v_{\rm p.iso} \propto R$, while for radii much larger than
the core radius ${\rho}_h \propto R^{-2}$ and $v_{\rm p.iso} \approx$
constant. Two free parameters describe the halo:
$R_{\rm core}$ and $v^h_{\rm max}$.

As mentioned is the previous section, a maximum disc fit assumes
a maximum contribution to the rotation curve by the luminous
components. For a system with only a disc this means in practice
that one scales up the disc rotation in the inner regions
as much as possible to be still consistent with the observations. 
In cases where there is a bulge too, because of its strong
mass concentration, the bulge contribution has to be maximized first
followed by the disc. For systems with a bulge the rotation
generally remains flat, going inwards to small radii. That implies that
if the bulge were not maximal, a dark halo with
very short core radius would be required. The maximum disc/bulge
adjustment procedure can generally be done with approximately a
5\% error in the bulge and disc rotation, which generates
a 10\% error in the M/L ratios. 

For the 12 galaxies maximum disc fits are presented in Fig.~2. 
In most cases the fits to the observations are excellent; even
certain small scale features in the photometry which are expressed
in the model rotation curve appear to be reflected in the observations. 
Nevertheless for two cases, NGC 5585 and DDO 154, there is a small
discrepancy at the outermost radii. It seems that the observed 
rotation drops while the model curves predict a continued rise.
Whether we are witnessing a steep end to the dark halo for
these two galaxies, or whether the observed
drop is an artifact remains to be investigated further. 
In Table~4 the fitting parameters are presented. 
Errors of the mass-to-light ratios are estimated at 20\%.
That is a combination of the errors of the amount of light, 
caused by the various 
uncertainties of distance, extinction, and colour corrections effects,
and the error generated by the fitting method (Sect.~10.3).
The determined core radius
is generally comparable in size to the radius of the last measured
rotation point. That is a direct consequence of the maximization
procedure of the disc which leads to a minimal amount of 
dark halo matter in the inner regions.   

\begin{figure}
\resizebox{\hsize}{!}{\includegraphics{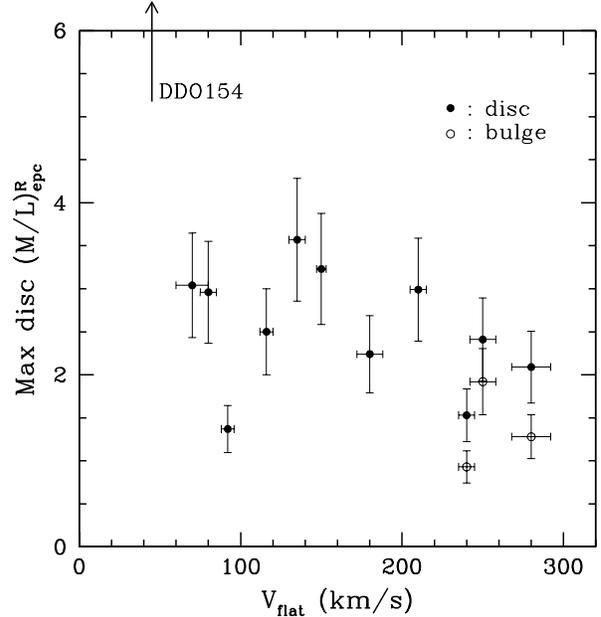}}
\caption{The M/L ratios, corrected for absorption and population
effects, following from the maximum disc fits. Different galaxies
are represented along the x-axis by their observed rotation
at the flat, outer part of the rotation curve. There appears
to be a trend in the sense that the more massive galaxies have
a smaller M/L ratio.}
\end{figure}

The determined M/L ratios corrected for absorption and population
effects are given as a function of the observed rotation at the
flat, outer part of the rotation curve in Fig.~3. 
It can be noticed that the corrected M/L ratios of the discs have
a range of at least a factor of two and a half. Minimum to 
maximum go from approximately 1.4 to 4 and to 9.2 for DDO154. There appears to be a
trend in the sense that the more massive galaxies have a smaller
corrected M/L ratio. Such an appreciable range has to be explained
by different initial mass functions for the galaxies, while 
evidence so far suggest that the IMF is more or less universal
for normal galaxies. One has to keep in mind, however, that the
maximum disc hypothesis is just a hypothesis. In a more practical sense
one can consider the determined M/L ratios as an upper limit. 
The smallest M/L ratio could then be the M/L ratio which applies for
all galaxies. In this respect it is interesting to note that
the (M/L)$_{\rm epc}$ ratios of the bulges are situated at the lower
range of the values. An equal M/L ratio for all discs and bulges 
for all galaxies given by, for example, the smallest maximum disc
M/L ratio would then imply that bulges remain close to the
maximum contribution to the rotation in the inner regions. 
A comprehensive analysis of these matters is made in Sect. 9. 

In Table~4 mass-to-light ratios are also given as observed,
so with no corrections. As can be noticed, then there is a large,
unrealistic, range of values. Consequently it is imperative that
absorption and population corrections are made. 

\begin{figure*}
\centering
\includegraphics[width=16cm]{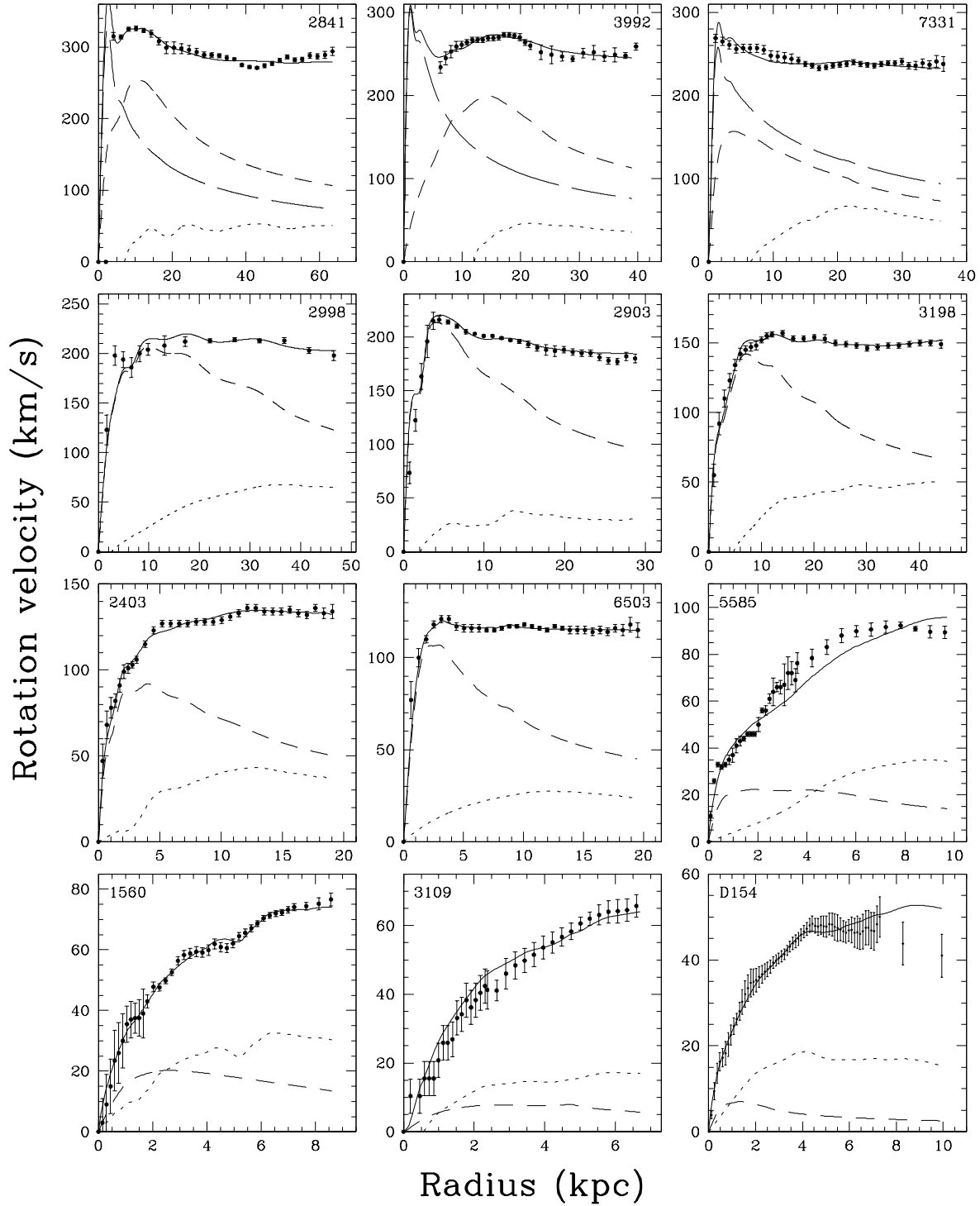}
\caption{MOND-like fits to the observed rotation curves. 
Here MOND-like means that there are two free parameters:
the M/L ratio of the disc (and bulge, if present) and the
MOND scale $a_0$. For most galaxies the quality of the fit
is excellent and comparable to that of the maximum disc case.
However, for three galaxies, NGC~5585, NGC~3109, and DDO~154
the fits slightly deviate from the data.}
\end{figure*}

\section{MOND fits}
\subsection{Basics}
As an alternative to dark matter, the flatness of rotation
curves may be explained by a different law of gravity which
prevails in the outer regions of galaxies. A rather successful
such proposal is Milgrom's (1983) modified Newtonian dynamics or 
MOND. Here the idea is that below a certain acceleration threshold
($a_0$) the effective gravitational acceleration approaches
$\sqrt{a_0 g_N}$ where $g_N$ is the usual Newtonian acceleration.
This modification yields asymptotically flat rotation curves 
of spiral galaxies and a luminosity -- rotation velocity relationship
of the observed form, $L \propto v^4$, the Tully-Fisher relation.
MOND is able to explain in considerable detail
the actual rotation curves of galaxies (BBS, Sanders 1996, 
Sanders \& Verheijen 1998, McGaugh \& de Blok 1998). In principle
$a_0$ should have one universal value and should not be allowed
as a free parameter. In that sense MOND rotation curves are 
predictions and not fits (Milgrom 1988).

\subsection{Two parameter MOND-like fits}
In first instance, however, MOND will be considered here
as a fitting procedure with $a_0$ and M/L as two free parameters,
and not as a phenomenological alternative to Newtonian gravity. 
As mentioned in the introduction, it is then compatible with the
other three fitting procedures employed in this paper, who also
use two free parameters. For the remainder, the two parameter MOND
fit will be referred to as MOND-like fit, to avoid any confusion.
To make this fit, in practice, a least squares fit to the observed
rotation curve is made by  having an acceleration 
$g_{\rm mond}$ equal to 
\begin{equation}
g_{\rm mond} = \frac{g_N}{\sqrt{2}} \left[
1 + \sqrt{ 1 + 4 \left( \frac{a_0}{g_N} \right)^2 }
\right]^{1/2},
\end{equation}
where the Newtonian acceleration $g_N$ is calculated from the mass
distribution in the standard way. The circular velocity is then
given as usual by $v_{\rm rot} = \sqrt{R g_{\rm mond}}$.

In Fig.~4 the MOND-like fits for the sample of galaxies are
presented while fit parameters are listed in Table~5. No large discrepancies
can be detected between the data and model rotation curves. For most
galaxies and certainly for the more massive, the fits are impressive.
But for three cases there are small yet noticeable deviations. For 
NGC 5585 the MOND-like fit gives too low rotations at intermediate
radii. A better representation can certainly be achieved by using
dark matter for a maximum disc situation (Sect.~5, Fig.~2), 
and especially for a slightly sub maximum disc case (Sect.~9, Fig.~9).
Thus the blame for this misfit cannot be put straightforwardly on an
incorrectly determined rotation curve. NGC 3109 has in the inner 
regions rotations which are slightly but systematically below
the fit, but considering the errors this is not a real problem.
More problematic is the obvious misfit in the outer regions
of DDO 154. For the other three fitting procedures who use specified
dark halo radial density laws there is the same misfit. Yet
an explanation for the decreasing rotation curve can then always
be found by a sudden end to the dark matter distribution.
In case of MOND that is not possible: there is no dark matter and
consequently the discrepant fit then poses a more serious problem. 

For the larger galaxies MOND-like requires the discs to be maximal.
This results in M/L ratios (Table~5) which are close to the values
found when using the maximum disc constraint. Consequently also
for MOND-like there appears to be a considerable range of M/L ratios,
both for the absorption and population corrected and for the
uncorrected M/L values. As to the likeliness of this one is
referred to the discussions in the previous section. 

\subsection{The value of $a_0$}

Figure~5 displays the fitted value of $a_0$ versus the rotation
of the galaxy at the flat, outer part of the rotation curve.
There is a considerable scatter, which remains if only the galaxies
with well determined Cepheid distances are considered. If the extreme
upper and lower points are ignored there might even be a trend such
that the more massive galaxies have a larger value of $a_0$ as was 
already tentatively reported by Lake (1989). 
A comparison can be made with the values determined by Randriamampandy \&
Carignan (2014) for a fixed (M/L)$_{3.6\mu}$ value. There appears
to be a comparable distribution of $a_0$ values; the scatter 
is comparable and there seems to be the same slight trend 
with galaxy mass. It thus appears that
the MOND-like fitting procedure also needs two free parameters. 
As such it can then give a good description of the observed rotations
of almost all of the galaxies in the sample. The fact that it is
possible to reproduce the observed rotation curves without
invoking dark matter inevitably points to a fundamental relation
between the baryonic matter distribution and the observed
large scale kinematics, whatever that might mean.  

{From} the perspective of Mondian philosophy, where one universal value
of $a_0$ is required, the data in Fig.~5 are not encouraging.
For the same analysis made by BBS, $a_0$ values are distributed
more closely around a constant, with one exception for NGC 2841,
and there is not a trend with mass of the galaxy. At that time
the distances to the objects were less accurate and moderate
adjustments to the individual distances made it plausible
that the rotation curves could be explained by MOND with a 
universal $a_0$ value determined at 1.21~10$^{-8}$ \cms\ by BBS.
Since then this number has been generally used by various
authors in a number of studies. 

In reality, however, there appears to be a reasonable uncertainty
concerning the universal value. Sanders \& Verheijen (1998) give 
MOND fits to the rotation curves of 30 spiral galaxies in the UMA
cluster which they assume to be at 15.5 Mpc. The preferred value of
$a_0$ with this adopted distance is equal to the BBS value. But, after
a Cepheid based re-calibration of the T-F relation, Tully \& Pierce
(2000) argue that the distance to the UMA cluster should be taken
at 18.6 Mpc. As discussed by BPRS the preferred value for $a_0$ then has
to be decreased to 0.9~10$^{-8}$ \cms. That is also the preferred
value of $a_0$ from MOND fits to rotation curves of a sample of 
nearby dwarf galaxies with distances taken primarily from group
membership (Swaters et al. 2010). 
Gentile et al. (2011) find an average essentially equal
to the orignal BBS value, but Randriamampandry \& Carignan
determine an average $a_0$ of 1.13 $\pm$ 0.50 10$^{-8}$ \cms. 
So, considering MOND as a real
alternative to dark matter, what universal value should one adopt? 

In the present analysis of which the sample largely overlaps
that of BBS, there is clearly no indication that $a_0$ should
be decreased to 0.9~10$^{-8}$ \cms. If one bears in mind the slight
trend of a decreasing $a_0$ for lower masses for eight galaxies,
as noted above, it might be plausible that certainly the dwarf galaxies
of Swaters et al. (2010) lead to a lower determination of $a_0$. 
Anyway, in first instance we take $a_0$ to be equal to 1.3~10$^{-8}$
\cms\ which is the median of the data in Fig.~5 and explore the 
consequences. 

\begin{figure}
\resizebox{\hsize}{!}{\includegraphics{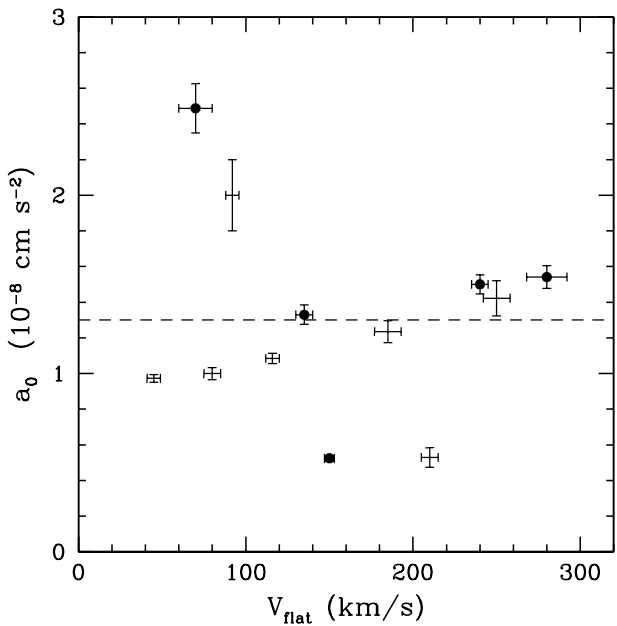}}
\caption{The value of the $a_0$ parameter following from the
MOND-like fits to the rotation curves of the sample.
The data appear to be uncorrelated and to have a considerable
scatter. Galaxies with Cepheid distances have been indicated
by a dotted point, the dashed line is at the median value.}
\end{figure}

\subsection{True MOND fits}
The fits to the rotation curves of the sample have been done again
but now for a fixed value of $a_0$ equal to the median of the sample.
It appears that for the eight galaxies belonging to the trend subsample,
for which $a_0$ in Fig.~5 ranges between 0.9 and 1.6~10$^{-8}$ \cms,
one can in general make a reasonable fit, having only M/L as a free
parameter. There are in some cases small systematic differences, but 
these could be explained by inherent systematic uncertainties associated
with the derivation of the rotation curve from the observed velocity
field. Even the most massive galaxy, NGC 2841, now at a distance of
14.1 Mpc, seems to be consistent with MOND.

\setlength{\unitlength}{1cm}
\begin{figure*}
\begin{minipage}{11.4cm}
\begin{picture}(11.4,10.9)
\resizebox{11.4cm}{!}{\includegraphics{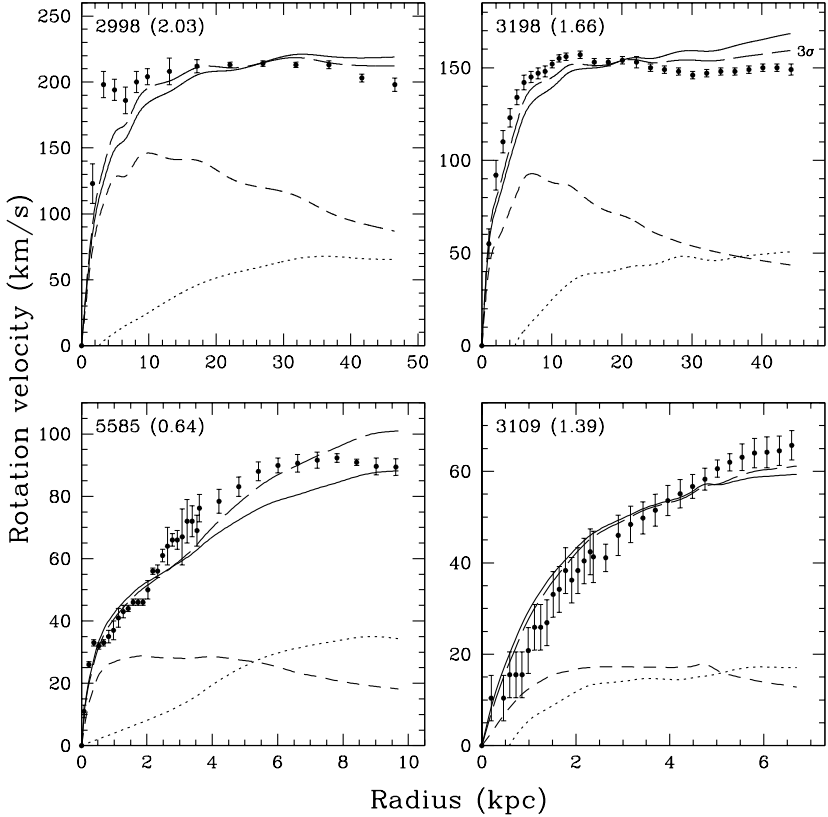}}
\end{picture}
\end{minipage}
\hfill
\begin{minipage}{5.7cm}
\begin{picture}(5.7,5.1)
\end{picture}
\caption[]{
One parameter MOND fits to the rotation curves of four
galaxies for which the two parameter fit generates
the most deviating $a_0$ value; using a fixed $a_0$
of 1.3~10$^{-8}$ \cms.
The full drawn line is the best fit,  while the long dashed line indicates
the fit for a $3\sigma$ more favourable
distance. The short dashed and dotted line are the disc and
gas contributions respectively. 
In principle, MOND should be able
to explain the rotation curves for all galaxies with
a universal value of $a_0$; which is obviously not possible.
Between brackets just after the galaxy name is the fitted value of $(M/L)^R_{\rm obs}$
using the nominal distance. 
}
\end{minipage}
\end{figure*}

The four outlyers in Fig.~5 have been considered in more detail.
MOND fits to the rotation curves of these galaxies using 
$a_0$ = 1.3~10$^{-8}$ \cms\ are presented in Fig.~6. Obviously,
in all four cases there is a clear and systematic deviation from
the observed data. For NGC 2998 and 3198 the fit in the inner regions
is too low by 10 to 20 \kmss and the fit in the outer regions is too
high by the same amount. For NGC 5585 and NGC 3109 the deviation is in
the opposite sense, the fit is too high in the inner regions and too
low in the outer regions. Considering the relations involved it is
clear that MOND prefers a smaller distance to the first two galaxies
and a larger distance to the latter two. Therefore, to assess the
seriousness of the misfit it has been investigated how far the
fit can be improved by taking a more favourable distance. 

\subsection{MOND?}
As a limit to what is possible the three sigma deviation has been
taken from the nominal distance quoted in Table~1. This three sigma
needs some explanation for the individual cases.
At first NGC 3198 which has a (metallicity corrected) distance
of 13.8 Mpc. The random 1$\sigma$ error amounts to 0.5 Mpc but the
systematic error for an individual galaxy as determined by
Freedman et al. (2001) is not immediately obvious. On page 54 of 
that paper, second equation, the systematic error is given
as the quadratic sum of the contributing errors of (1) the 
zeropoint LMC PL relation, (2) the metallicity,
(3) the photometric zero point and (4) of aperture correction
for bias, crowding, background etc. The quoted numbers for errors
(1) to (4) are 5\%, 3.5\%, 3.5\%, and 0 to 5\% (we use 2.5\%) respectively. 
Adding in quadrature gives the systematic error for an individual
galaxy of 7.5\% and thus 1.03 Mpc for NGC 3198. The total, 
systematic plus random error for NGC 3198 then amounts to 1.14 Mpc
(1$\sigma$) and 3.42 Mpc (3$\sigma$) and the more favourable
distance at the 3$\sigma$ level is then 10.38 Mpc. This is a conservative
error since the WMAP measurements (Spergel et al. 2007) 
indicate that the Hubble constant
determined by Freedman et al. is in reality more accurate than quoted. 
For NGC 3109 the Cepheid distance of 1.36
$\pm$ 0.10 Mpc is used (Musella et al. 1997), but as noted before,
recent observations by Soszynski et al. (2006) suggest the Cepheid
distance to be at 1.30 $\pm$ 0.04 Mpc. Considering these numbers
the plus three sigma limit has been put at 1.6 Mpc. NGC 2998
has no determined Cepheid distance. Yet the galaxy is relatively
far away at a Hubble distance of 67.4 Mpc. 
The error in the distance is taken as the quadratic
sum of the errors of the Hubble constant (for 1$\sigma$ 5 \kms\
per Mpc) and a possible deviation from a regular Hubble flow
(250 \kms\ at 1$\sigma$). A three sigma more favourable distance
is then at 50.6 Mpc instead of the nominal 67.4 Mpc. 
For NGC 5585
there is a Hubble distance of 6.2 Mpc and there is a distance of
8.7 Mpc from an analysis of the brightest stars (Drozdovsky \&
Karachentsev 2000) which the authors qualify as ``somewhat uncertain''.
Anyway, for the present this 8.7 Mpc is, admittedly somewhat 
arbitrary, considered as the $3\sigma$ upper distance limit. 

For the more favourable limiting distances the MOND fits using
the median value of $a_0$ are given by the long dashed lines in Fig.~6.
Compared with the fits using the nominal distance for the 
galaxies NGC 2998, 3198, and 3109 there is some improvement.
As can be noticed,
the rotation curve of NGC 2998 is sampled by a limited number of 
points because the galaxy is distant. The deviation of the fit
from the data remains systematic and significant but only for a handful
data points. For NGC 3198 the difference between the fit and data
is systematic and large for the majority of a large number of 
independent rotational data points although the fit can be made
less deviant by taking an outmost smaller distance. NGC 3109 also shows
a systematic deviation over the whole radial extent. Yet the errors 
on the rotational points are considerable which leads to a difference
which is significant but not extreme. For NGC 5585 the fit can be improved
considerably by taking an acceptably larger distance; the correspondence
between the fit and data is then comparable to the fit when taking 
$a_0$ as a free parameter (Fig.~4). That fit, however, was already
noted to be not so good, certainly when compared with the fit
using DM. For the adopted median value of 1.3~10$^{-8}$ \cms\ for $a_0$  one
can then conclude that for MOND there is a small problem for NGC 5585,
a problem for NGC 2998 and NGC 3109 and a serious problem
for NGC 3198. In second instance a universal value for $a_0$ of 0.9~10$^{-8}$ \cms\
has been investigated. The MOND fit then improves for NGC 3198 and 
as demonstrated by BPRS, putting the galaxy $2\sigma$ closer makes
the fit marginally consistent with the data. Also the MOND fit for 
NGC 2998 improves markedly, but the fit for NGC 3109 and NGC 5585
becomes considerably worse. However, for this value of $a_0$ NGC 2841
gives serious and significant problems (see BPRS). 

When $a_0$ is allowed as a free parameter a large range of values
is found. That is essentially the reason for the subsequent finding 
that it is not possible to make acceptable MOND fits for all 
galaxies for any value of $a_0$. This does not add to the
credibility of the theory.

\section{Adiabatic Contraction}
Galactic dark matter halos formed in the universe by cumulating matter cannot
preserve their original density distribution. The baryonic mass 
component within a halo will collapse mainly dissipational and
in doing so pulls dissipationless dark matter to the centre. 
Thus when finally a galaxy has formed its dark matter halo has 
a different radial distribution compared to distributions predicted
by pure CDM cosmological simulations. A prescription is needed to 
relate the DM profiles of the present sample of galaxies to those
of the pre collapsed galaxies. 

\subsection{Standard Adiabatic Contraction}
Such prescriptions or mechanisms have been designed and investigated
in several studies. The original specific and comprehensive work on
this matter is by Blumenthal et al. (1986). They consider the fate
of a halo with an original homogeneous mixture of dissipationless
DM and baryonic matter. Such a halo contracts when the baryons fall in
dissipatively. When the baryonic mass fraction ($F$) is small and
for the simplifying approximation that the orbits of dissipationless halo 
particles are circular, the quantity $rM(r)$ is an invariant, where
$M(r)$ is the total mass within radius $r$. For this situation the
orbits of halo particles essentially change adiabatically and the
mechanism has been named Adiabatic Contraction (AC).

Assume that the initial spherically symmetric cumulative mass distribution
$M_i(R)$ at initial radius $R$ is a mixture of collisionless DM $(M_{i,{\rm dm}}(R))$ and fraction
$F$ of baryonic matter, $F = M_{i,{\rm bar}}/M_i$. The baryons cool
and fall into a final galaxy cumulative mass distribution
$M_{\rm gal}(r)$ at final radius $r$. Then the adiabatic invariant implies that

\begin{equation}
r\left[ M_{\rm gal}(r) + M_f(r) \right] = R M_i(R)
\end{equation}
with the continuity condition of
\begin{equation}
M_f(r) = (1 - F) M_i(R)
\end{equation}
Here $M_f(r)$ is the final DM cumulative mass at final radius $r$
contracted from the initial cumulative distribution 
$M_{i,{\rm dm}}(R)$. In practice, for an
initial distribution $M_i$, for every initial radius $R$ one can solve
equations (9) and (10) to get the final radius $r$. Converting
to rotational velocities, the final rotation $v_f(r)$ is then 
given by

\begin{equation}
v_f(r) = \sqrt{ \frac{(1-F)M_i(R)}{r}} = \sqrt{ \frac{M_{i,{\rm dm}}}{r} } 
= v_{i,{\rm dm}}(R)\sqrt{\frac{R}{r}}
\end{equation}
This AC prescription is not straightforward; the final distribution
and rotation curve has to be calculated for every radius separately using
given but not always simple cumulative mass distributions. From now
on we shall refer to the mechanism described above as the 
standard AC model. 

Blumenthal et al. explore this standard AC for various cored
dark matter and disk distributions. In general an original pseudo isothermal
mass distribution changes into a distribution which is much
more centrally concentrated. Therefore assuming both, that the 
standard AC is correct and original halos are pseudo isothermal, it is
unlikely that present day halos have strictly pseudo isothermal profiles. 

\subsection{Modified AC}
The matter of AC on itself has been re-investigated by
Gnedin et al. (2004). They perform and analyse numerical simulations
mainly of clusters, including a.o. gas cooling, dissipation, and
star formation. In general the result is that standard AC overpredicts
the concentration of DM in the inner regions of the clusters. Gnedin
et al. propose a modified AC prescription, which they tested to give
a good representation of the density profiles of simulated clusters. 
The proposed modification is to use the quantity $rM(\bar{r})$ as the
invariant instead of $rM(r)$. Here $\bar{r}$ is the orbit averaged
radius of an isotropic distribution which can be approximated 
fairly well by
\begin{equation}
\bar{r} = A r^w
\end{equation}
over a wide range of radii with $A \approx 0.85$ and $w \approx 0.8$.
Equations (9), (10), and (11) then change into
\begin{equation}
r\left[ M_{\rm gal}(\bar{r}) + M_f(\bar{r}) \right] = R M_i(\bar{R})
\end{equation}
\begin{equation}
M_f(\bar{r}) = (1 - F) M_i(\bar{R})
\end{equation}
and
\begin{equation}
v_f(\bar{r}) = v_{i,{\rm dm}}(\bar{R}) \sqrt{ \frac{\bar{R}}{\bar{r}} }
\end{equation}
For the single galactic halo they investigated the correspondence
between the simulation and the modified prescription was
less satisfactory. 

\subsection{Testing AC}
Later on Choi et al. (2006) also investigated these matters.
Amongst others 3D N-body simulations have been carried out of galactic
DM halos with an isotropic velocity distribution. Various galaxy
formation scenarios and DM halo concentrations have been considered
representing the different results of cosmological simulations.
Choi et al. find that for NFW halos the standard AC prescription gives,
on average, halo rotations at $R = 2.2 h_{\rm disc}$ which are
6\% too large. 
For Gnedin's modified AC prescription with $A = 0.85$ and $w = 0.8$
the correspondence between the halo rotations of theory and of simulations
is satisfactory, in general. On the other hand it is not possible
to reconcile the structure of the emerging cored halos with predictions
of any theory. Using the standard AC the halo rotations at 
$R = 2.2 h_{\rm disc}$ are about 20\% too large compared to that of
the simulated halos. Considering masses this gives errors larger
than 40\%, a huge number. In fact, this means that the results
of Blumenthal et al. concerning pseudo isothermal cored halo distributions
are not correct, in retrospect. 

The AC theory works better for initially more concentrated
halos. As Choi et al. argue, this can be explained by the distribution
of the orbits. For isotropic halos, in order to be less concentrated,
a larger fraction of more radial orbits is needed. Therefore
the assumptions underlying standard AC become less adequate for
less concentrated dark halos. Consequently one may expect a worse
correspondence between the predictions of standard AC and real
contraction for shallower halo distributions. This also explains
the results of Jesseit et al. (2002) that standard AC works well
for Hernquist (1990) halos which are even more concentrated than
NFW halos. 

\subsection{The AC which has been used}
In the next section rotation curve decompositions will be considered
for NFW halos. Such halo shapes are predicted by cosmological
CDM formation scenarios. Subsequent galaxy formation will therefore
make the original density profile more concentrated and an AC
mechanism is needed to relate the original profile to the present
day halo distribution and rotation curve. In accordance with the 
discussion above the modified AC prescription will be used. 
In Sect. 9, following the results of Sect. 8, cored pseudo isothermal
halos are considered. The resulting halo rotation curves from the
rotation curve decomposition procedure have been adiabatically 
de-contracted in order to generate pre baryonic collapse density
profiles. But how should one do this de-contraction when the
standard procedure produces rotations which are off by 20\% and
even the modified AC is not correct? As suggested by Gnedin et al.'s
analysis, AC can be made gentler by using the $\bar{r}$ 
of Eq.~(12). We have investigated if AC can be made even gentler
by changing the values of $A$ and $w$. It appears that for
$w = 0.5$ instead of 0.8 the AC rotations of the halo at 
$R = 2.2h$ can be lowered by the amount to bring them in agreement
with the simulation results of Choi et al. Therefore, to do the
adiabatic de-contraction of cored halos the invariant 
$rM(\bar{r})$ is used with $\bar{r} = 0.85\; r^{0.5}$; and it is
checked if indeed the de-contracted halo rotation is in agreement
with a cored distribution. This procedure is somewhat artificial,
but until more sophisticated AC methods become available, it is
the best one can do.  

\begin{figure*}
\centering
\includegraphics[width=16cm]{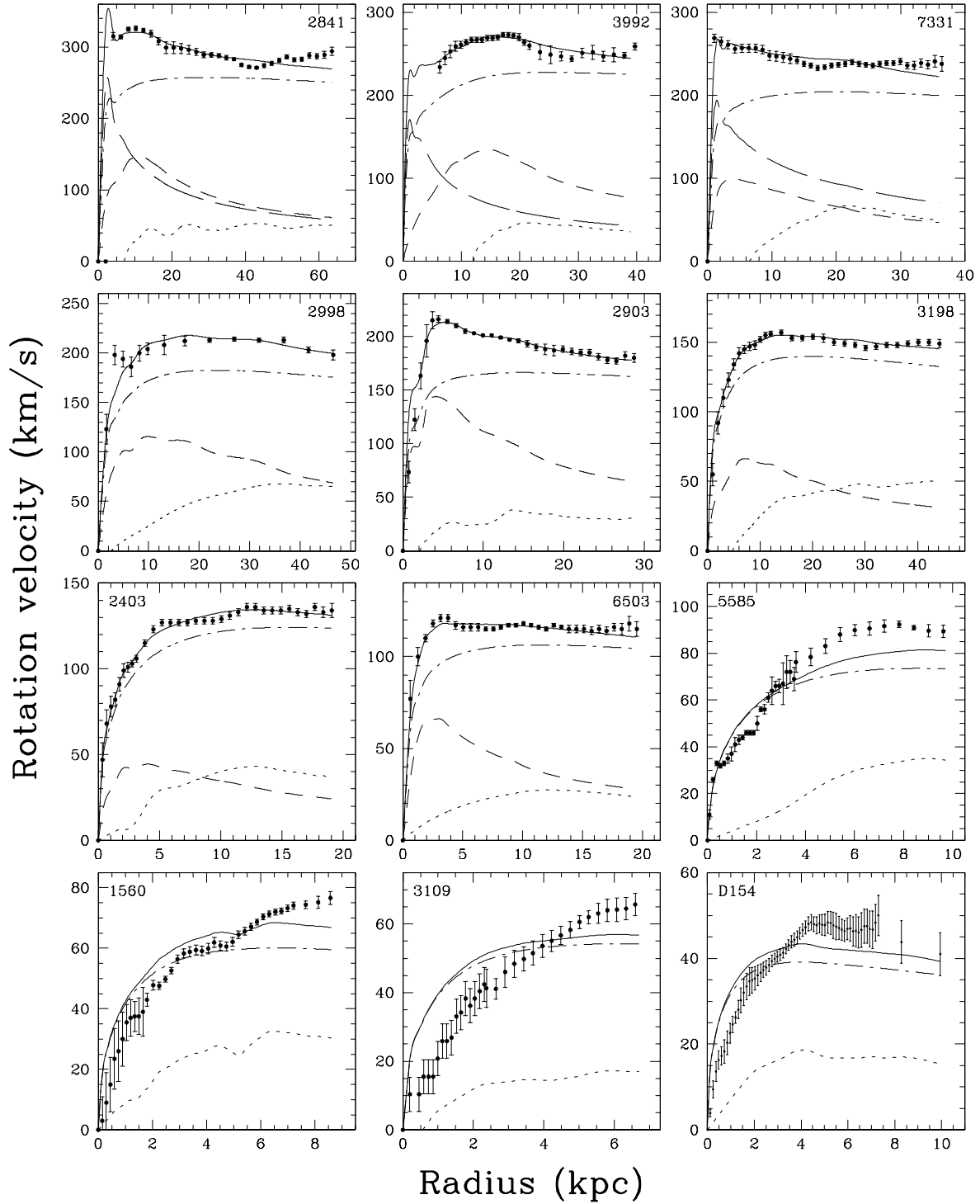}
\caption{
Fits to the observed rotation curves for NFW-CDM$\Lambda$ dark
halos including the appropriate process of adiabatic contraction.
For the 8 most massive galaxies an adequate representation of
the observed rotation can be achieved, though in some cases
only for unrealistically small M/L ratios. For the 4 least luminous
galaxies an NFW halo is clearly inconsistent with the observations
even for a most favourable situation, as presented, where the 
luminous contribution has diverged to zero.
}
\end{figure*}

\setlength{\unitlength}{1cm}
\begin{figure*}
\begin{minipage}{11.4cm}
\begin{picture}(11.4,10.9)
\resizebox{11.4cm}{!}{\includegraphics{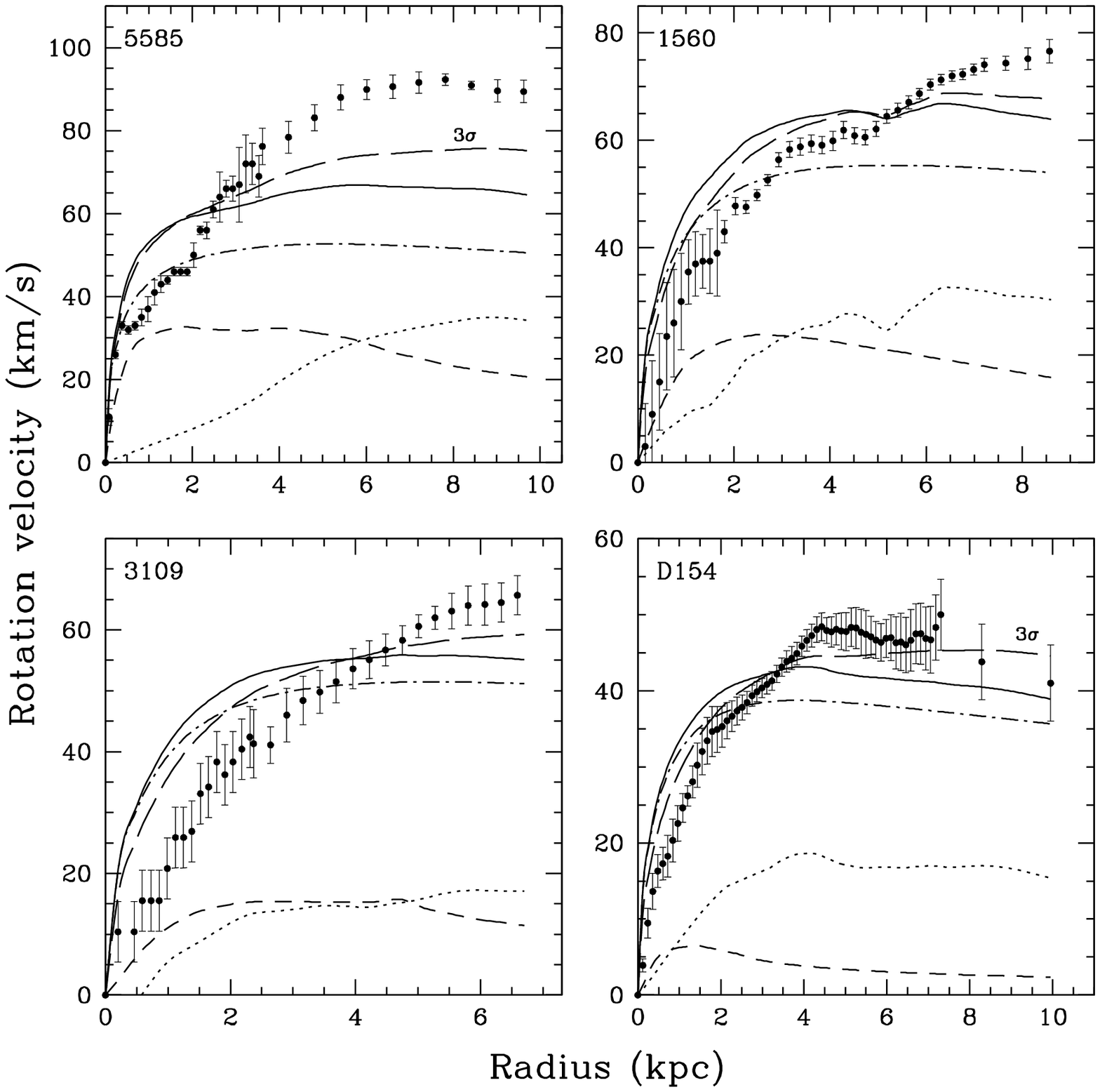}}
\end{picture}
\end{minipage}
\hfill
\begin{minipage}{5.7cm}
\begin{picture}(5.7,5.1)
\end{picture}
\caption[]{
NFW-CDM$\Lambda$-AC fits to the rotation curves of the
four least massive galaxies, using a realistic yet moderate
M/L ratio for the disc. The discrepancy between fit and data then
becomes larger compared to the situation without luminous matter,
as in Fig.~7.
Full drawn lines indicate the best fit, the dotted, dashed and dash-dot lines
are for the gas, disc, and dark halo contributions respectively. 
In addition the fits for $3\sigma$ less concentrated dark halos
are given by the long dashed lines. Even for such an extreme deviation
from the average properties the mismatch remains.
}
\end{minipage}
\end{figure*}

\section{NFW-CDM$\Lambda$-AC fits}
\subsection{A one parameter NFW halo}
For an NFW halo the density distribution takes the form
\begin{equation}
{\rho}_{\rm NFW} =  \frac{{\rho}_i}{(R/R_s)(1 + R/R_s)^2}
\end{equation}
where $R_s$ is a characteristic radius and ${\rho}_i$ is related
to the density of the universe at the time of collapse. The
rotation curve $v_{\rm NFW}$ following from this distribution is
\begin{equation}
v_{\rm NFW} = 2.15\; v_{\rm max} \sqrt{ \frac{R_s}{R} {\rm ln} 
\left( \frac{R}{R_s} + 1 \right) - \frac{R_s}{R + R_s} }
\end{equation}
where the maximum rotation $v_{\rm max}$ is reached at a radius of 
$2.16 R_s$. When a certain cosmology is chosen the structural
parameter becomes related to the total mass of the halo. 
Following Bottema (2002) we assume the current concordance model
for the cosmology: a low density CDM universe with flat geometry,
called CDM$\Lambda$ with ${\Omega}_0 = 0.25$, $\Lambda = 0.75$, and a
Hubble constant of 75 \kmss per Megaparsec. For that cosmology
Navarro et al. (1997) relate the parameter $M_{200}$ to 
$v_{\rm max}$ in their Fig.~7. Some manipulation with equations leads
to the following relation between $R_s$ and $v_{\rm max}$
\begin{equation}
{{R_s}\over{[{\rm kpc}]}} = 0.0127 \left( {v_{\rm max}}\over
{[{\rm km}\;{\rm s}^{-1} ]} \right)^{1.37}
\end{equation}
such that when Eqs.~(17) and (18) are combined there is
only one free parameter $(v_{\rm max})$ left for the dark halo. 
One may relate the present parameters to
that of the concentration parameter $c$, often appearing in the
literature by
\begin{equation}
c = 55.74 \left({ {v_{\rm max}}\over{[{\rm km}\;{\rm s}^{-1} ]} } 
\right)^{-0.2933}
\end{equation}
of which the value will be given as well.

One way to proceed when making NFW halo decompositions, is to maintain
two free halo parameters and do a least squares fit to determine
e.g. $c$ and $V_{200}$ (de Blok et al. 2001; Navarro 1998). For each 
galaxy a data point is generated in the $c, V_{200}$ plane and the
distribution of points in this plane can be compared with
the distribution generated by numerical simulations for
various cosmologies. For an observed rotation curve which
is close to solid body, however, problems will occur. An NFW halo
has $v_{\rm halo} \propto \sqrt{R}$ in the inner regions. When
such a functionality is fitted to $v_{\rm obs} \propto R$ the fit
parameters get forced into an extreme region of parameter space:
$c$ and $V_{200}$ diverge to a small and large value
respectively. One then has a situation where a small galaxy
is embedded in a gigantic dark halo. The diverged structure of this
halo is barely related to the galaxy under investigation and a
comparison with numerically generated dark halos is not
appropriate. In practice one circumvents this divergence
by setting artificial bounds to e.g. $c$ (see de Blok et al. 2001).

Since the cosmology has now become better established,
constraints on NFW halos like those given by
Eq.~(18) can be used. This is fortunate because then
the degeneracy can be relieved between disc mass and dark halo 
contribution for a fit to an observed rotation curve. In addition,
a diverging fit to an inner solid body rotation curve
for small galaxies does not appear any more.
For the remainder of this section NFW fits will be made to 
rotation curves using two parameters (M/L ratio and $v_{\rm max}$)
for systems without bulges and three parameters ( + M/L ratio of
the bulge) for galaxies with bulges. 
In practice the constraining relation (18) is an average and
cosmological simulations display a certain scatter around the relation. 
If one investigates only one galaxy this scatter has to be taken
into account by considering the probability that 
the halo properties of that particular galaxy deviate
from the average. But, for a collection of systems, as presently,
an appearing systematic deviation from the average relation 
is sufficient to demonstrate an inconsistency. 

\subsection{Fitting the rotation curves (with AC)}
Cosmology provides the constraint which is badly needed, but
not without a trade-off, namely the necessity to apply an AC procedure.
This procedure has been applied to the halos of the sample of
galaxies. For a certain M/L ratio of disc (and bulge) the radial
cumulative mass profile of the baryonic component has been calculated
from the photometric and \hi density radial profiles. An NFW halo
rotation curve parameterized by the value of $v_{\rm max}$
is then contracted to a new rotation curve which cannot be described
by an analytical function any more. Furthermore, the shape of the
curve is different for each different M/L ratio and obviously a 
straightforward least squares fitting method cannot be applied.
Instead, the only way to proceed is by generating ${\chi}^2$
difference values in the plane of the parameters M/L and $v_{\rm max}$
and searching this plane for the minimum ${\chi}^2$ value. When a bulge
is included the plane changes into a cube, of course, with the
additional parameter the M/L ratio of the bulge. These procedures
have all been performed and the final fits to the rotation curves
are presented in Figure~7 and numerical values in Table~6.
A few remarks before we investigate the fits. The value of the halo's
maximum rotation $v_{\rm max}$ given in Table~6 applies to the original
NFW halo including the mixed in baryonic fraction. One can
calculate the maximum rotation of the DM component only by
$v_{\rm max}^{\rm h, dm} = \sqrt{1 - F_{200}} \; v_{\rm max}^{\rm h, tot}$
where $F_{200} = M_{\rm bar}/M_{200}$ also given in Table~6.
The values of $R_c$ and $c$ follow directly from that of 
$v_{\rm max}$ by Eq's (18) and (19). Quoted errors have been
inferred by taking the same fractional error as generated by the standard
Marquard least squares fitting of a non contracted NFW halo.
For NGC 7331 the disc contribution has to be small, but is poorly
constrained. It has been fixed at the low value suggested by the
observed stellar velocity dispersions (Bottema 1999).

As can be seen in Fig.~7, for the more massive galaxies the fits are
generally very acceptable and not worse than the maximum disc
or MOND-like fits. There is, however, a significant difference when
comparing the NFW fits with the previous ones. The disc is forced
to a contribution which is substantially lower than the maximum
possibility, in line with the conclusions of Dutton et al. (2005).
This results in M/L ratios which are lower than
the maximum disc and MOND values. Considering the galaxies with
observed stellar velocity dispersions (excl N7331) the maximum
contribution of the disc rotation to the total rotation is 
54\%, 42\%, and 55\% for NGC 2998, 3198, and 6503 respectively. Such
values are at, or just below the lower limit inferred from the
observed dispersions. 

Comparing the fitting results with those obtained without the
AC procedure one can note the following. For the more massive and
certainly the more concentrated galaxies the effect of AC on the
halo RC is more pronounced in the sense that the halos become
more concentrated. Which is obvious, of course. For the smaller
galaxies and the more so for those with only a minimal disc the
AC effect is small. The gas mass distribution is relatively extended
and has as such only a limited influence on the contraction. 
In general, when AC has been included the contribution of the 
luminous components becomes even smaller than it already was for
NFW halos without AC. The bulge rotations are, on average,
an extra 20\% lower and the disc rotational contributions
$\sim$ 10\% lower. 

\subsection{Disagreement for small galaxies}
As can be readily noticed, for the less massive galaxies with
$v_{\rm max} \la$ 100 \kms, the fits are not compatible with the data. 
There is no way the deviations can be explained by uncertainties
or errors in the observed rotation. What can be noted as well
is that the fits force the M/L ratio to zero. If instead a reasonable
M/L ratio is adopted, the discrepancy between fit and data becomes
even worse. This has been investigated by making extra NFW-AC
fits for the four least massive galaxies assuming an M/L ratio
which is modest in any way. For that we assume (M/L)$_{\rm epc} = 1.0$
following a.o. from stellar velocity dispersion observations and which
is roughly half the value for a maximum disc contribution.
The result is displayed in Fig.~8 and as can be seen, the 
correspondence between best fit and the data becomes even worse
compared with the M/L = 0 case. Two effects contribute to this
deterioration. At first the addition of a stellar contribution, but
secondly, because of that, the halo density becomes slightly more
concentrated by an increased AC effect. 

\subsection{Shallower NFW halos}
The possibility has been explored that by some chance these
four galaxies have formed in dark halos which were extremely shallow.
Considering the Navarro et al. (1997) halos, which have been used
in the present fitting procedure, Bottema (2002) derives a compromise
value for the $1\sigma$ scatter in ${\rm ln}(c)$ of 0.25
based on the analyses of Bullock et al. (2001) and Jing (2000).
This value of the scatter appears to be exactly equal to the value found by 
Macci\`o et al. (2008) for relaxed halos resulting from later, more
sophisticated simulations. That scatter value can thus be used with some confidence. 
It is easy to demonstrate that a scatter in ln($c$) translates in exactly
the same scatter in ln($R_{\rm s}$) and so in a scatter to the 
coefficient 0.0127 in Eq.~(18). A shallower halo by $1\sigma$ then gives
a coefficient of 0.0163 and by $3\sigma$ of 0.027; for the same
$v_{\rm max}$ the associated $R_{\rm s}$ is then increased. 
For a $3\sigma$ less concentrated halo fitting results
are displayed in Fig.~8 as well. It can
be noticed that the correspondence between data and best fit improves,
but the discrepancy remains large. 
Values of the concentration parameter $c$ for the nominal fits are
given in Table~6 and amount to an average of 17.5 for the four least luminous
galaxies. For $3\sigma$ less concentrated halos the value of $c$ is then
a factor 2.12 less or approximately 8. 
Making a two parameter NFW halo fit to the rotation curves
can generate acceptable fits, but concentration parameters then
decrease to a value below 4 while the core radius parameter $R_s$
ends up above 12 kpc and the mass-to-light ratio tends towards zero. 
This is a situation of a small galaxy embedded in a giant DM halo,
as discussed above. 

Using the Millennium Simulation Neto et al. (2007) establish that
NFW halos appear to be shallower than the halos investigated by 
Navarro et al. (1997). A relation is found for NFW halos of
$c = 5.26\; (M_{200}/[10^{14} h^{-1} M_{\sun}])^{-0.10}$
for $M_{200} \ga 10^{12} M_{\sun}$ which is
compatible with the relation
$c = 5.6\; (M_{200}/[10^{14} h^{-1} M_{\sun}])^{-0.098}$
for $M_{200} \ga 10^{10} M_{\sun}$, so for smaller halos, 
as derived by Macci\`o et al. (2007).
The relation by Neto et al. can be converted in a
$R_{\rm s}$ versus $v_{\rm max}$ relation as in Eq.~(18) using
$h = 0.75$:
\begin{equation}
\frac{R_s}{[{\rm kpc}]} = 0.0197 \left( \frac{v_{\rm max}}
{[{\rm km}\;{\rm s}^{-1} ]} \right)^{1.37}
\end{equation}
thus having the same exponent as in Eq.~(18) but a larger factor
of 0.0197 instead of 0.0127, expressing that the halos are indeed
shallower. Then halos of the shape of Neto et al. (2007) are equal
to the NFW halos of Eq.~(18) but shallower at a deviation of
1.7$\sigma$. If such fits were to be plotted in Fig.~8, these would
then be between the displayed nominal and $3\sigma$ fits. 
In fact, fits of Neto et al. halos, being $1\sigma$ more shallow than
their average would nearly coincide with Navarro et al. (1997) halos
which are $3\sigma$ shallower. As can been seen, such fits are
very deviant from the observations; the chance of finding four galaxies
which are shallower by $1\sigma$ from the average amounts to 0.00063.
It appears that
the value of the concentration slightly depends on the exact cosmology
(Macci\`o et al. 2008; Ludlow et al. 2014) and now seems to swing
backwards from the shallow Neto et al. (2007) determination to 
a larger concentration. Anyhow, the fitting possibilities
allowed by different cosmologies and simulation techniques should be
sufficiently well illustrated in Fig.~8.

The standard CDM halo
formation scenarios predict halos with an NFW profile on all scales.
Consequently that scenario is not consistent with observed
rotation curves of less massive galaxies, as mentioned already in
the introduction, and demonstrated here rather dramatically.
It is even in conflict
with RC analyses for intermediate galaxies because of the predicted
too low disc contributions. A more general and extended discussion
of these matters is deferred to Sect.~10.

\begin{figure*}
\centering
\includegraphics[width=16cm]{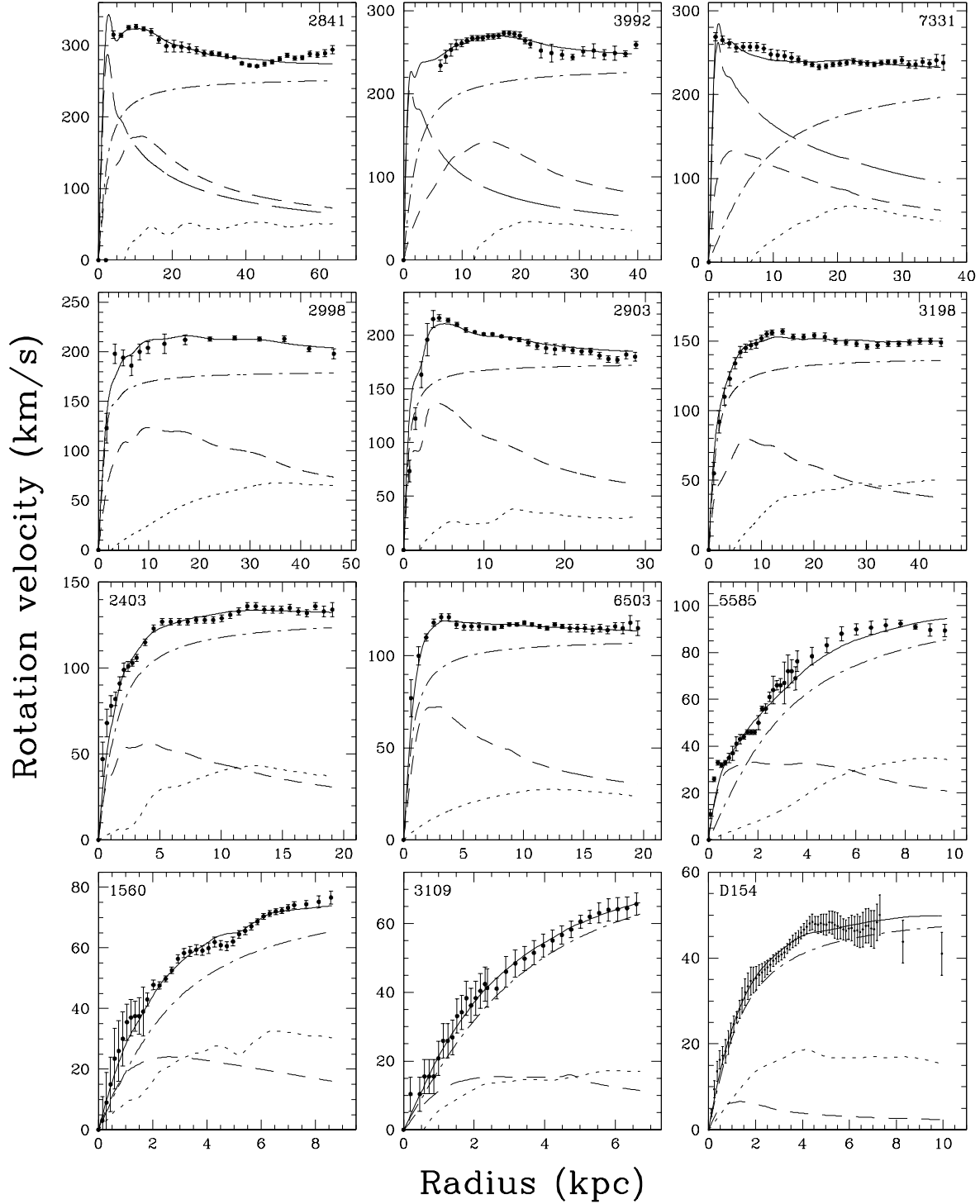}
\caption{
Fits to the observed rotation curves for one universal
value of the M/L ratio ($(M/L)^R_{\rm epc} = 1.0$) and
pseudo isothermal dark halo. There is an excellent
agreement between the model and data. For this fit
procedure the rotational contribution of the luminous
components is set a priori by the amount of light
and the colour; in general discs appear to be sub
maximal and bulges maximal. 
}
\end{figure*}

\begin{figure*}
\centering
\includegraphics[width=16cm]{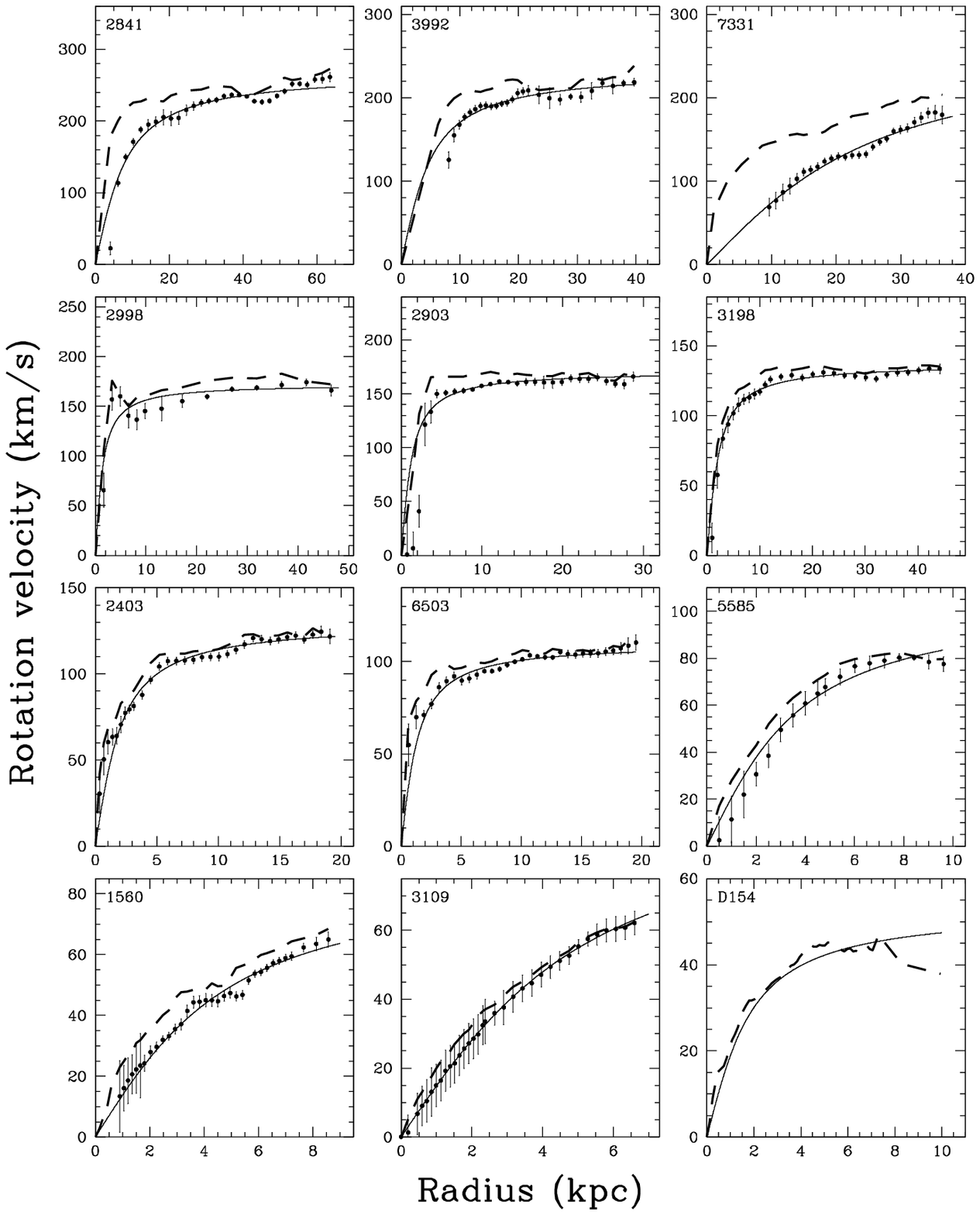}
\caption{
Dark matter rotation curves which result after subtraction
of the rotation of the luminous components, all with equal
universal M/L ratio, and of the rotation of the gas. 
The dashed line gives the result derived directly from the
observations, the points represent the adiabatically
de-contracted dark halo rotation, resampled to the original
radial positions and with appropriate errors. To these points
a fit is made of a pseudo isothermal halo given by the
full drawn line.
For DDO154 the data points have been omitted to avoid confusion;
for that case the de-contraction is very small. 
}
\end{figure*}

\section{Equal mass-to-light ratio fits}
Under normal galactic conditions, the IMF
seems to be universal or at least barely dependent on metallicity
or star formation rate (Kroupa 2001, Bell \& de Jong 2001). 
Because the M/L ratio is directly
related to the IMF one expects this ratio to be universal too, at
least when properly corrected for extinction and population effects.
Therefore rotation curve decompositions will be made assuming
an equal (M/L)$_{\rm epc}$ for all the galaxies. 

\subsection{The value of the M/L ratio}
What value should one use? It certainly has to be smaller than the
smallest maximum disc and maximum bulge values displayed in Fig.~3,
or else some rotation curves will rise above the observations. 
The appropriate value for any disc can, in principle, be derived
from measured stellar velocity dispersions. Unfortunately such a 
derivation is not straightforward. At first, because the observed
stellar velocity dispersions come with a considerable error; the
M/L ratio is proportional to the square of the dispersion and hence
M/L even has a larger uncertainty. Secondly, in order to derive
the mass-to-light ratio, assumptions have to be made for the scalelength
to thickness ratio $(h/z_0)$ of a disc, the ratio of vertical to radial
velocity dispersion, and absorption corrections. For four galaxies
of the sample values are given in Table~7, all derived for appropriate
disc parameters and conditions. Excluding NGC 7331, which has the most
uncertain value because its disc light is so dominated by the bulge,
an average (M/L)$_{\rm epc} = 1.04\; \pm \sim 0.3$ is found. 
Considering this number and the above-mentioned requirement to have
an (M/L)$_{\rm epc}$ below the lowest maximum disc/bulge numbers, 
a plausible value for the universal (M/L)$_{\rm epc}$ of 1.0 has been
chosen. It could equally well be 0.9 or even 1.1 or 1.2 yet for now
the consequences will be explored for a reasonable equal value for 
all discs and bulges of the sample of galaxies. 

\subsection{Fitting the rotation curves}
Using a pseudo isothermal halo and (M/L)$_{\rm epc} = 1.0$ for all
components, the resulting fits are presented in Fig.~9 and Table~8.
The agreement between model curves and the data points is excellent
in general. The exception being the outermost velocities of DDO 154, 
which don't fit in any case. In general the discs are sub maximal
which is a logical consequence of an assumed M/L ratio resulting from
observed stellar velocity dispersions (Bottema 1993; Martinsson et al. 2013).
However, for the cases where the disc contribution should be larger
because a specific disc feature is clearly expressed in the observed
rotation curve, the disc contribution is indeed larger, for
example for NGC 2903 and NGC 5585. Moreover, the bulges give
a dominant contribution to the rotation, which they should do in
order to keep the rotation flat near the centre. It can be concluded
that one universal M/L ratio provides the possibility to make excellent
fits to observed high quality and extended rotation curves. Contrary to 
maximum disc fits there is a physical basis for the adopted M/L ratio
and it can be calculated for every galaxy a priory.
Considering the results just mentioned, the argumentation can be
turned around. If there is indeed a universal (M/L)$_{\rm epc}$ ratio
what can its value be by only investigating the rotation curve fitting?
Making it larger than $\sim 1.1$ lifts the RCs above the observations
for NGC 2841, 7331, and 5585. On the other hand, making it smaller
than $\sim 0.9$ the fit for NGC 2903 gets considerably worse.
Thus, in retrospect, the choice of (M/L)$_{\rm epc} = 1.0$ was appropriate,
though admittedly, we iterated once on this matter and increased
the number from 0.9 to 1.0.

\subsection{Pure dark matter}
Following the findings and discussion above a more fundamental approach
can be taken to derive the contribution of the dark halo.
Assuming that one knows for every galaxy what the contribution of
the luminous component is, taking (M/L)$_{\rm epc} = 1.0$, one can simply
subtract the luminous contribution from the observed rotation curve and retain
the rotation of the dark halo only. This has been done for the sample
and the resulting halo rotation curves are given in Fig.~10 by the dashed lines. 
In this way a nice impression is obtained of pure dark matter. 
Fits of any functionality can be made to these dark halo rotation curves.
For the specific pseudo isothermal halo functionality of Eq.~(6)
nearly identical results would have been obtained as for the combined
fit performed to the total rotation curves as tabulated in Table~8.
To relate the data to the original, pre galaxy formation dark halos
the rotation curves of the halo have to be adiabatically de-contracted. Using
the scheme described in Sect.~7 this has been done under the assumption
that the original dark halos are pseudo isothermal. The result is given
in Fig.~10 too, where the de-contracted halo RCs have been re-sampled
to the original radial positions and the errors have been copied from
the contracted halo rotation curves. Fits of a pseudo isothermal functionality
have been made to these rotational data points, being very acceptable
(except for DDO 154 again) and resulting fit parameters $R_c$ and
$v_{\rm max}$ are given in Table~8.

\begin{figure}
\resizebox{\hsize}{!}{
\includegraphics{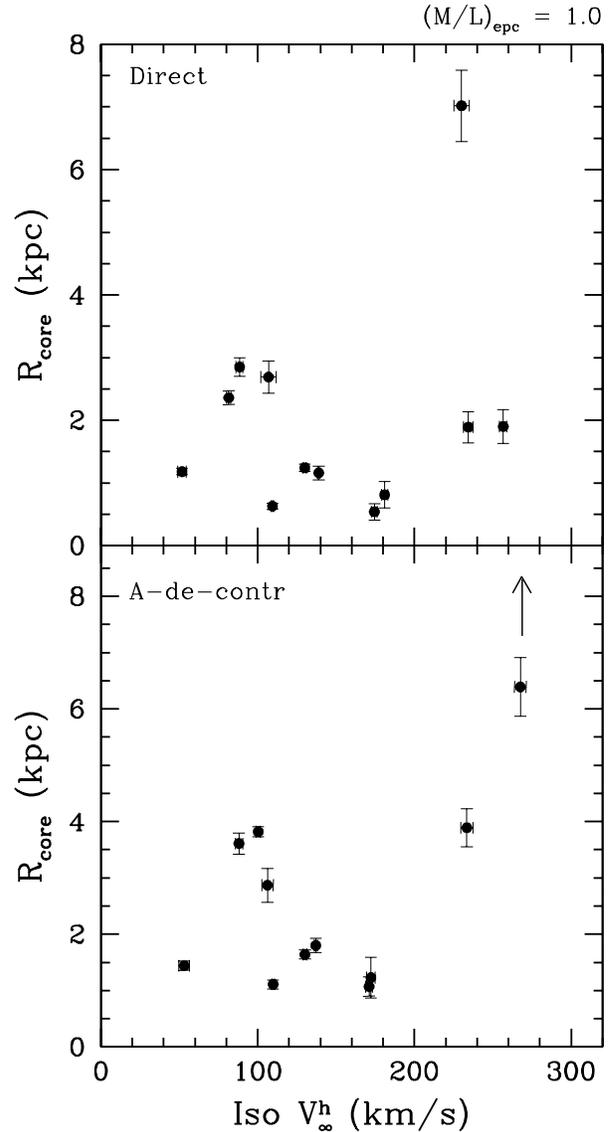}}
\caption{
Core radius versus maximum rotation of the pseudo isothermal halo
fit to the dark halo rotation curves in case of the adopted
universal M/L ratio for the luminous components.
{\it Top:} for the composite rotation curve fit, directly to
the data as in Fig.~9. {\it Bottom:} for the fit to the dark
halo which results after subtraction of the baryonic rotation
and subsequent adiabatic de-contraction, as in Fig.~10.
There is a moderate range of core radii, but no trend
or correlation seems apparent. In the bottom plot the off-scale
value for NGC 7331 is indicated by the arrow.
}
\end{figure}

The derived halo core radii have been plotted versus halo maximum
rotation in Fig.~11, both for the direct fit to the total observed 
RC and for
the fit to the adiabatically de-contracted dark halos on itself. In general
$R_{\rm core} \la$ 4 kpc and in first instance no clear correlation
or trend seems visible. The process of adiabatic de-contraction makes
the core radii larger, as expected, by some 30 to 100\%, but the
general appearance in Fig.~11 does not change significantly. 
For every galaxy the value of the core radius is rather specific and
changes by more than 50\% quickly give a disagreement between model and 
observed rotation curves. But this is only so for the employed fixed disc 
and bulge contribution at (M/L)$_{\rm epc} = 1.0$. It might
be investigated if changes of this M/L ratio for each galaxy, within the boundaries
of the uncertainties, can stretch the possible range of fitted core radii,
and if an equal value for all core radii for all galaxies is possible. 
Results of such an investigation will be presented in a future paper.

\subsection{Other than pseudo isothermal halo densities}
The investigation of the dark matter distribution for a universal
M/L ratio has concentrated, so far, on a pseudo isothermal radial
functionality and can as such give a consistent explanation of
the observations. But how about other DM radial functionalities and
specifically that of an NFW halo? To investigate this for the sample
of galaxies the halo RCs obtained after subtracting the baryonic
components with (M/L)$_{\rm epc} = 1.0$ have been adiabatically
de-contracted using $w = 0.8$, appropriate for NFW halos. Then
there is a first problem for the three galaxies with bulges. For these
the process of adiabatic de-contraction with $w = 0.8$ is so severe
that for the inner halves of the halo RCs non realistic 
decontractions are found. For the intermediate galaxies, the decontraction
works fine and in general a satisfactory fit of a two parameter
NFW rotation curve (Eq.~17) can be made to the data. But there is
nothing new. For example, let us consider the galaxy NGC 3198. The
pseudo isothermal fitting procedure to the halo RC gives a good
fit to the data over the range between $2R_c$ and $25 R_c$.
Over that range of radii the value of dln$\rho$/dln$r$ of the halo
goes from -1.5 to -2. For the NFW fit procedure the fit is acceptable from
$R = 0.2 R_s$ to $3 R_s$ and then dln$\rho$/dln$r$ ranges between
-1 to -2.5. So considering the radial functionality of the dark halo
density there is in fact not much difference between a pseudo isothermal
and an NFW halo, and both can over a specific range of $R/R_{c,s}$
describe the observations. 

A difference between pseudo isothermal and NFW can only be noticed there
where both differ. That is in the outer regions where for the
NFW profile dln$\rho$/dln$r$ goes to -3 while for a pseudo isothermal
profile it cannot go below -2. Translated in RCs that means that
in order to distinguish NFW from pseudo isothermal a maximum rotation and
decline beyond has, or has not, to be observed. Inspection of Fig.~10
shows that (except for DDO154) the RCs of the halos continue to rise
and consequently there is no evidence, at large radii, that an NFW
profile is to be favoured over a pseudo isothermal profile. 

Also in the inner regions the profiles differ. An NFW halo does not
have a dln$\rho$/dln$r > -1$ (the cusp) while for small radii
a pseudo isothermal halo has $-1 <$ dln$\rho$/dln$r$ $< 0$ (the core).
For example the halo of the galaxy NGC 1560 has after adiabatic decontraction
a dln$\rho$/dln$r$ ranging from $\sim -0.1$ to -1.7 which 
cannot be reconciled with the RC of an NFW halo. The rotational
data points of the halo are simply too close to solid body rotation
to be reconciled with $v_{\rm rot} \propto \sqrt{r}$ over the 
observed range of radii. Also for two other small galaxies, 
NGC 5585 and NGC 3109, it is not possible to fit an NFW two parameter
RC to the halo data points. Therefore it can be concluded that
under the assumption of a universal (M/L)$_{\rm epc}$ value of 1.0
an NFW radial density profile is not consistent with a considerable
fraction of the galactic dark halos. 

Burkert halos (Burkert 1995) combine the constant density
(isothermal) inner core with an $R^{-3}$ (NFW) functionality
at large radii. Fits of the rotational functionality of such
halos have been made to the A-de-Contracted halos of the sample
as presented in Fig.~10. It appears that for the galaxies more
massive than NGC 5585 the fits differ systematically from the
data; in the inner and outer regions data lie above the fit
while in the central regions it is the other way around. 
This is because a Burkert halo has a maximum rotation while
for a good fit a radial section with nearly flat rotation
is needed. For the small galaxies Burkert works well. This is
easily explained because for these systems the rotation still
rises, and the curve is fitted with the inner, constant density
part of the Burkert halo. If also the M/L ratio were left as
a free parameter, the fit would converge to a maximum disc situation.
The then remaining halo is forced into the constant density
inner section of the Burkert functionality and a better
fit is achieved.

\begin{figure}
\resizebox{\hsize}{!}{
\includegraphics{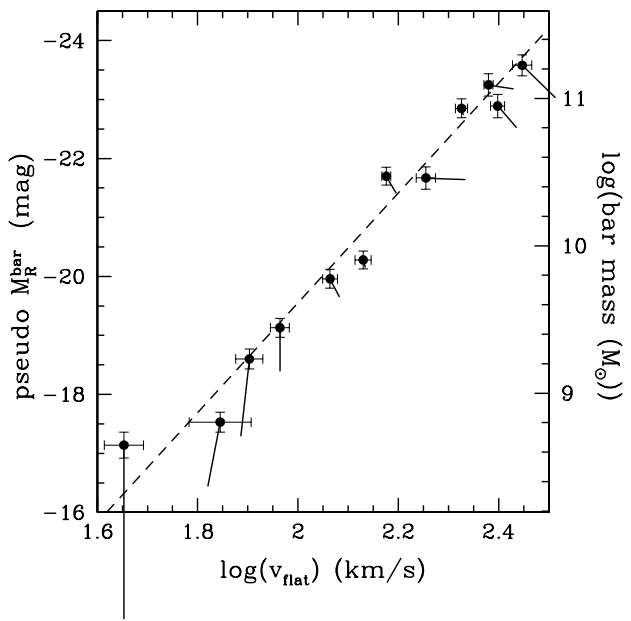}}
\caption{
A baryonic TF-like relation for the sample of galaxies
using $(M/L)^R_{\rm epc} = 1.0$. Mass or associated luminosity
is given as a function of velocity at the observed, flat outer part of the
rotation curve. There is an exquisite linear relation over a 
mass range spanning more than $2\frac{1}{2}$ orders of magnitude.
The fit of a straight line gives a slope of 
$3.7 \pm 0.2$ (dashed).
In addition a more classic TF relation, with no population correction and
no gas mass included, is given by displacing the 
data points to the ends of the attached solid lines. 
}
\end{figure}

\subsection{Tully-Fisher-like relations}
To further investigate any physical correlations a few
Tully-Fisher like relations have been put together. At first,
in Fig.~12, the total baryonic mass is given as a function
of the rotational velocity at the flat part of the rotation curve.
The errors on the baryonic mass have been taken at 20\%, which is
estimated to be the combined uncertainty of absorption and
population corrections and distance (see Sect.~10.3 for a discussion).
To determine the error on
$v_{\rm flat}$ an inspection by eye is made of the rotation curve in
the outer regions. Fig.~12 displays an astonishingly good linear
relation over $2\frac{1}{2}$ orders of magnitude in galaxy mass.
A fit of a straight line to the data points has been made,
taking the errors in both directions into account (Reed 1989)
giving a slope of 3.7 $\pm$ 0.2 and reduced ${\chi}^2$ of 2.5.

A second TF-like relation is given in Fig.~13 where the total
baryonic mass has been plotted versus the maximum rotation of the
pseudo isothermal halo resulting from the direct (non A-de-C) fit
to the individual rotation curves. Again there is a good linear 
relation with a slope of 4.3 $\pm$ 0.4 and reduced ${\chi}^2$ of 4.8
thus being slightly steeper than that of the previous TF-like
relation. This is explained easily by the fact that for the least
luminous galaxies the fitted halo rotation curve continues to rise
beyond the last measured rotational points. The asymptotic halo maximum
rotation is then relatively larger compared to that of the more massive
galaxies. As is obvious, the scatter for this relation (Fig.~13) is larger
compared with that of the previous one in Fig.~12. This may be explained by 
the inclusion of the rotational effect of the baryonic component
into the observed value of $v_{\rm flat}$ on the x-axis of Fig.~12.
That creates a moderate
correlation between the ordinate and abscissa making the
scatter smaller compared to that of the completely uncorrelated axes
in Fig.~13. In both TF-like relations NGC 3198 is slightly deviating
in the sense that it is too massive (or bright) for its rotation. 
This is likely caused by the fact that NGC 3198 contains
a large amount of gas compared to its light (see Table~1) such that
for this galaxy more mass in the form of gas has fallen into its halo.  

\begin{figure}
\resizebox{\hsize}{!}{
\includegraphics{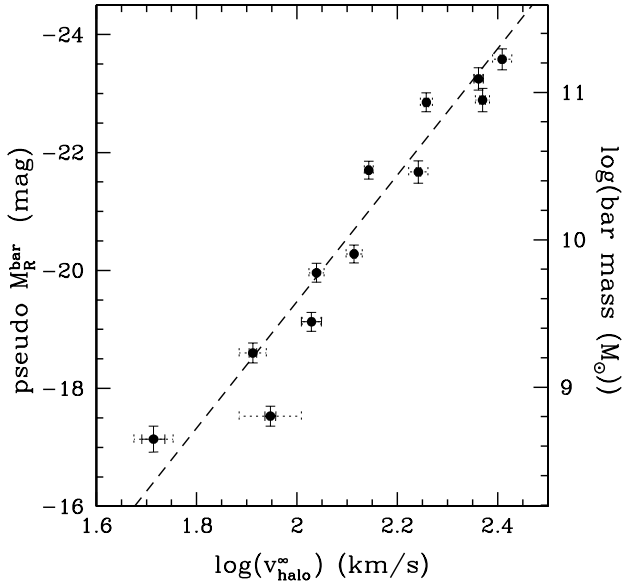}}
\caption{
As Fig.~12, but now baryonic mass is plotted versus
the maximum rotation of a pseudo isothermal dark halo resulting
from the fit to the observed rotation curves as in Fig.~9.
Errorbars in the x-direction give the formal error by the
drawn lines and the error estimate by eye, equal to that in
Fig.~12, by the dashed lines. A fit of a straight line to 
the data (dashed line) has a slope of $4.3 \pm 0.4$.
}
\end{figure}

The original Tully-Fisher relation (Tully \& Fisher 1977) is a relation
between the absolute magnitude of a galaxy, corrected for extinction,
and the line width of the \hi profile, corrected for inclination and
internal broadening. Depending on the way the observations are being
treated and the way the corrections are applied, generally a different
relation is found. The two relations presented here, in a natural way,
contain much less intricacies compared to the classical relation. 
At first, the rotation is used instead of the line width. Furthermore,
a population correction is applied to the luminous component and gas
mass is added to obtain a total baryonic mass. This should automatically
remove two deviating features or specific ``kinks'' associated with
the classical TF relation (see Noordermeer \& Verheijen 2007).
Usually the brightest galaxies appear to be somewhat too faint for
their rotation. This kink at the bright end can be largely removed
by taking the rotations at the flat asymptotic part of the
rotation curve and by using near infra-red colours instead of optical
colour to diminish absorption and population effects. A second 
deviation at the faint end, below $v_{\rm rot} \sim$ 90 \kms,
exists. This kink can be
eliminated by using the baryonic TF relation (McGaugh et al. 2000) thus
including the gas mass which generally constitutes a larger fraction
for the fainter spiral galaxies. 

To investigate these matters for the present sample, also a more classic
TF relation has been put together. No population correction is applied, 
only the extinction corrections, which for the systems with bulges
have been applied to the total light using the inclinations derived
from the \hi velocity field. The maximum observed rotation is taken
and gas mass has not been included. Then a correlation between the
R-band luminosity and rotation is constructed
and is displayed in Fig.~12 by a displacement of the datapoints.
Instead of a linear relation a curved
relation is present as expected following the discussion in
the preceding paragraph.

In general, the main conclusion of studies dealing with TF relations
is that there must be a fundamental relation between the mass of
the dark halo and the mass of the baryons that have fallen in. 
For the present study the same conclusion may be drawn based on the
tight linear relations that have been derived and are presented
in Figs~12 and 13. Yet for the assumed universal M/L ratio and
ensuing dark halo properties that conclusion can be checked more
directly. The total baryonic mass is known for each galaxy. In addition
the rotation curves of the adiabatically de-contracted halos, which
represent the pre formation halo structures, are given in Fig.~10. 
{From} that the cumulative mass of these halos until the last
measured data point can be calculated directly, which has been 
done actually. In a number of cases, however, the outermost rotational
points of the halos show some scatter. Therefore, to obtain a 
representative value for the velocity at the end of the
measurements (at $R_{\rm end}$) the rotation at the end of the fit
of the pseudo isothermal rotation curve to the data points is taken. 
The error on this value is estimated by eye.
Then in Fig.~14 the baryonic mass is plotted
versus the halo mass at $R_{\rm end}$. As can be seen, indeed,
there is a clear and rather linear relation between the two
mass parameters, now confirming directly the main suggestion
of TF studies. Inspection of Fig.~14 shows that it appears that the
more luminous galaxies contain a relatively larger fraction of
baryons. But, there is a caveat, however. Not the total halo mass is
presented but only the mass until the outermost measured rotational
point. So only conclusions can be drawn concerning masses until
that radius.

\begin{figure}
\resizebox{\hsize}{!}{
\includegraphics{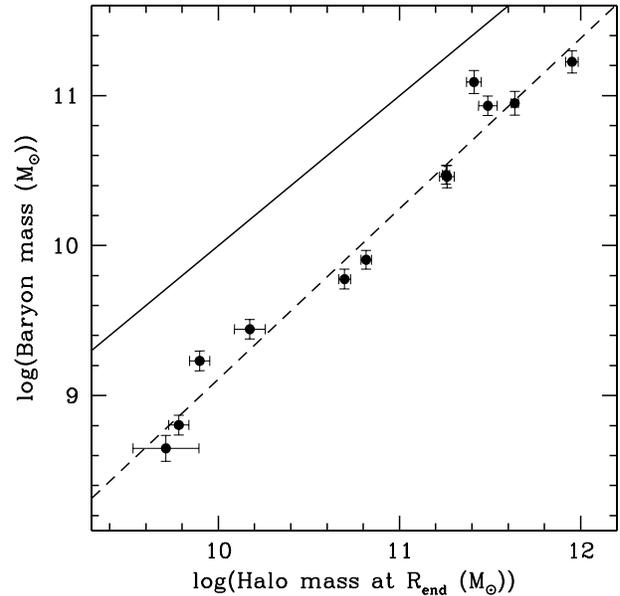}}
\caption{
The baryonic mass of the galaxies using the universal
$(M/L)^R_{\rm epc} = 1.0$, as a function of total mass of
the dark halo at the radius of the outermost measured rotation.
The apparent well established correlation confirms the notion
that there is a fundamental relation between the mass
of the dark halo and the mass of the baryons that have
fallen in. The full drawn line represents unity; the dashed line
is a linear fit to the data, having a slope of $1.14 \pm 0.07$.
It appears that more massive galaxies contain a larger
fraction of baryons; typically the ratio of DM to baryons
is $\sim$ 9 for small galaxies and $\sim$ 5 for large galaxies.
(Mind: data points of NGC 3198 and
NGC 2903 nearly coincide).
}
\end{figure}

\section{Discussion and conclusions}
Some general matters will be addressed and some items will be
discussed in more detail. Finally the main conclusions are
given point by point. 
\subsection{Comparing the four fitting procedures}
Four fitting procedures have been employed, in principle each
with two free parameters. Comparing these four can be done 
by inspecting figures 2, 4, 7, and 9, and judging the quality of
the fits. All fits have been achieved by minimizing the ${\chi}^2$
difference between model and data, but as discussed in Sect. 4, care
has to be taken when interpretating this
minimum value of ${\chi}^2$. An inspection by eye 
may yield additional and a more general insight. 

One can immediately notice that the NFW fit is by far the worst,
certainly for the least luminous galaxies. In that case for the other galaxies
the contribution of the luminous component is forced to low values. 
An NFW halo is simply too concentrated to be able to properly represent
the dark matter in a galaxy. 
For the other three procedures, the quality of the fits is
comparable. MOND-like generally converges to a maximum disc contribution
for the massive galaxies, but to a submaximum disc situation for
the least massive systems. That procedure has also a slight
problem fitting the rotation curves for NGC 5585 and NGC 3109. 
As noted in sections 5 and 6, a maximum disc contribution leads
to a M/L ratio, corrected for extinction and population effects,
which varies considerably from galaxy to galaxy with a general trend
to decrease with a factor of two going from the least to the most
massive systems. Such a change in M/L ratio is not to be favoured,
or at least it is difficult to explain.

It appears that the MOND-like fitting procedure generates a considerable
range of values for the free parameter $a_0$, and thus, like the
maximum disc or equal M/L procedure also two free parameters are
required to make the fits. One may wonder if there is a fundamental
difference between MOND-like and the other two. To assess this matter,
it should be considered what the fitting does in each case. MOND-like
as it has been applied here, keeping the distances fixed,
uses Eq.~(8) to scale up the Newtonian rotation. Instead the other fitting schemes
add the rotation of supposed dark matter to the Newtonian rotation. 
As such, all use a two parameter scaling of the already existing
rotation of the luminous matter. Yet, for MOND-like the scaling
parameters are both in the direction of the velocity, while for
fixed M/L values with a dark halo, one parameter scales in the velocity 
and the other in the radial direction. So there is a fundamental
difference and the adding of dark matter is a more flexible method.
But by no means, judging the fits of the present sample, 
one method is to be preferred over the other. 

The maximum disc procedure maximizes the amount of luminous matter
and minimizes the amount of dark matter. Hence the result is certainly
useful in constraining properties of galaxies. In the paper by
Van Albada \& Sancisi (1986) discussing the maximum disc principle
a number of arguments are presented favouring a maximum baryonic
contribution. However none of these arguments present any hard 
evidence. Observations of stellar velocity dispersions
(Bottema 1993; Kregel et al. 2005)
lead to a sub maximum contribution, yet, locally in the disc
the baryonic matter dominates. This observation in combination with
the adoption of a universal M/L ratio, provides the fitting
scheme described in Sect. 9 and in Fig.~9. The results of this
scheme are by no means worse then those of the maximum disc or
MOND-like fits, and in some cases even better. So being physically 
and observationally motivated, and providing a good representation
of the observed rotations, that procedure is preferred. 
Because of that a rather extensive analysis is made of the 
predicted dark matter properties following from the universal
M/L ratio method. Results from that analysis should be useful
in constraining dark halo and galaxy formation scenarios. 

\subsection{\hi scaling}
It appears that for spiral galaxies,
the ratio of total mass to \hi mass in the
outer regions is roughly constant (Bosma 1981; Carignan \& Beaulieu 1989;
Broeils 1992a, b). As a consequence it might be argued that the
observed extended flat rotation curves can be explained by
scaling up the gas rotational contribution, a procedure
called \hi scaling. These matters have been investigated in
some detail by Hoekstra et al. (2001), who showed that this
procedure works well to explain the observed shapes of the rotation
curves for a sample of 24 spiral galaxies of various morphological types. 
They found an average \hi scaling factor of 9; correcting for the presence
of helium, this implies a total gas scaling factor of approximately 6.5.
Hoekstra et al. found a systematic trend with luminosity,
lower luminosity galaxies require a lower gas scaling factor 
and vice versa. Later on these matters have been addressed by
Swaters (1999) for a sample of dwarf spiral galaxies and by 
Noordermeer (2006) for a sample of early type spirals. It appears
that the dwarfs require substantially smaller scaling factors 
than Hoekstra et al. found and that the early types, in many cases,
require substantially larger scale factors. 

For the sample in this paper the \hi scaling has also been investigated.
The results were rather disappointing. In three cases the fit
is bad in all respects. For the rest, generally, the global correspondence
between fit and data is reasonable, however in most cases there is
a misfit for details. There are two cases (NGC 2998 and NGC 6503) where
the global fit is good and the detailed fit reasonable, while only one case
(NGC 3198) gives an excellent fit. The total gas scale factor shows
a considerable range, from 4.7 to 22.5, seven galaxies have values
around 6.5, but five out of 12 have substantially larger values. 
Considering the discussion and findings above,
it is obvious that there is not a universal gas scaling
factor. The rotation curve fit then needs two parameters: M/L ratio
and gas scale, and as such does not improve on other fit procedures.

\subsection{Accuracy of the M/L ratios}
The uncertainties of the employed M/L ratios stem from a couple
of sources. At first the distance errors. For those determined
by the Cepheid method the error is typically $ \la$ 10\%; the
distances of the other galaxies are less certain. M/L scales inversely
linear with distance and so the error is passed on linearly.

Concerning the determination of the total amount of light one is
hampered by the uncertainties of the extinction and population corrections.
For the extinction corrections, a well motivated and well supported
method is used; dependent on the total
luminosity of a galaxy, as should be done for a sample with a large
range of masses (Tully et al. 1998). The internal absorption correction
of a face-on galaxy is less certain, but this correction is always
relatively small and consequently the error of the method is of less
influence. A reasonable estimate of the total extinction error is 0.12 
magnitudes. 

The population correction as such is rather well determined.
Bell \& de Jong (2001) have extensively tested the M/L
dependence on colour for several galaxy formation scenarios
and for several population synthesis codes. In all cases
a tight correlation is found for the optical passbands. Going
to the near infra-red, M/L becomes dependent on metallicity
and age too. Consequently for a sample with broad galaxy parameters
near infra-red colours should be avoided when restricting M/L ratios.
At present, the main error stems from the uncertainty of the
absorption corrected B-V colour. It is practically impossible to
derive this correction on B-V from basic principles for a patchy dust
distribution in combination with local star formation activity for
a spiral galaxy. 
Instead the empirical procedure of the RC3 is used which is
estimated to give, on average, an error of $\sim$ 15\% on the 
population correction. Adding the errors for distance, extinction,
and population correction in quadrature one ends up at a
total error slightly larger than 20\%. 
That implies approximately an error of 10\% for rotational velocities;
though it has to be kept in mind that for galaxies with uncertain
distances this error can be significantly larger.

\subsection{Where do the dark halos end?}
In this paper it is demonstrated that the
pseudo isothermal density functionality gives an excellent description
of the density distribution of dark halos. But, as can readily
be noticed, the $R^{-2}$ functionality at large radii generates
a mass which increases linearly with $R$. At some point the
halo has to end; meaning that the density at some point
has to deviate from pseudo isothermal.

As discussed in Sect. 9, for a fair number of galaxies
a density law according to Burkert (1995) is not in agreement with the
determined rotation of the dark halo. A Burkert halo has
an isothermal centre in combination with an $R^{-3}$ decrease at
large radii. Obviously such a density functionality drops off too
quickly and in reality a density $\propto R^{-2}$ over an extended
range is necessary. This conclusion agrees with the same finding
from an analysis of strong lensing observations for early
type galaxies (Koopmans et al. 2006).

Is there a sudden end to the dark halo or a smooth
transition to the ambient density of the local universe?
The former possibility seems to be suggested by the galaxy DDO 154.
For that system, dominated by gas, the observed rotation
drops nearly Keplerian beyond a radius of
5.5 kpc ($\sim 11h$), which indicates a sudden end
of the dark halo. However, such a drop is only observed
for one case. As such the evidence is not overwhelming.
Moreover, because the galaxy is not massive, the rotation
is small, and the velocity field is poorly resolved.
A confirmation of this drop is needed and more \hi rich
dwarf galaxies might be investigated for the presence of
this phenomenon.

\subsection{Cores generated by SIDM?}
The analysis of this paper adds to the evidence that the central
regions of DM halos have a central constant density core.
This contradicts the result of halo
formation calculations using collisionless DM which inevitable,
by violent relaxation, produces a central cusp with a density
proportional to $R^{-0.5 \rightarrow -1.5}$ (NFW, Moore et al. 1999,
Diemand et al. 2004). To solve this contradiction one can
invoke astrophysical processes to save CDM (see the introduction
for a discussion), or alternatively
explore other kinds of DM (Dav\'e et al. 2001, and references therein).
Anyway, looking at the available evidence to date, the constant
density core emerges triumphantly everywhere. An obvious and well
known astrophysical counterpart of such a structure is a globular
cluster. The formation of these is supposed to have taken place by
the collapse of giant gas clouds, primarily in the early universe. 
By solving the spherically symmetric, virialized gravothermal fluid
equations (Lynden-Bell \& Eggleton 1980) the formation and structure
of globular clusters can be explained. By the similarity of cored
halos and globulars one is then drawn to the suggestion that halos
may also have been formed by gas-like of self interactive processes. 

Such an idea of Self Interacting DM (SIDM) has been investigated
semi analytically (Spergel \& Steinhardt 2000; 
Ahn \& Shapiro 2005) and numerically
(Yoshida et al. 2000; Dav\'e et al. 2001; Col\'{\i}n et al. 2002).
Obviously the problem is more complicated than that of a collisionless
situation and results to date are rather preliminary. One clear and
non controversial finding is that SIDM can indeed produce halos
with cores, for an intermediate range of realistic DM particle
scattering cross sections. In a cosmological setting these cores can be
stable over a Hubble time. Controversy exists over whether the size
of the cores varies with mass of the halo or with cosmological
environment. It appears that SIDM generates cores with equal central
densities because this density is essentially set by
collisional physics. These densities are comparable to the observed
densities of halos of LSB and/or dwarf galaxies (Firmani et al. 2001),
ranging between 0.01 to 0.04 M$_{\sun}\; {\rm pc}^{-3}$.

At this stage a comparison can be made of the SIDM predictions with the
result of the rotation curve decompositions, especially those of
Sect. 9 for a universal M/L ratio. The core radii of the dark halos
derived from the decompositions appear to be rather independent
of galaxy and halo mass (Fig.~11), if any, the core radii
of the smallest and most LSB galaxies are a factor of two larger
than those of the intermediate and larger sized galaxies.  
The simulations of Dav\'e et al. and Col\'{\i}n et al. seem to indicate that
more massive halos have larger core radii, although the resolutions
are still poor and results preliminary. Central halo densities
of the four smallest galaxies are in the range of 0.01 to 0.03 
M$_{\sun}\; {\rm pc}^{-3}$, in agreement with the findings of 
Firmani et al. and with the numerical SIDM calculations. For the 
intermediate galaxies without bulges, however, the central halo 
density is found to be a factor of ten larger. This contradicts 
the SIDM results at first sight. On the other hand, the galaxies
investigated here are rather isolated and cores made up of self
interacting particles might have undergone various stages of collapse,
because they lack heat input by continual infall of DM. In this respect
it is interesting to note that the galaxies with bulges seem to have
a lower central core density again. Hypothetically that could suggest
a kind of bulge-DM interaction or replacement during the initial
stages of galaxy formation. Anyway, the derived structure of dark
halos from the present analysis of a sample of galaxies with a
broad range of sizes and properties, can set clear limitations to
any possible alternative to collisionless DM. 

\subsection{Conclusions}
\begin{enumerate}
\item
{For a maximum disc situation with a pseudo isothermal dark halo
excellent fits can be made to the observed rotation curves.
Yet an uncomfortably large range of M/L ratios is required.}
\item
{A MOND fitting scheme with the two free parameters,
$a_0$ and mass-to-light ratio, can give
a good representation of the observed rotations.}
\item
{MOND with only one free parameter, the M/L ratio of the luminous
component, cannot explain the rotation curves for all galaxies
simultaneously, not even when the distances to the galaxies
are stretched to their limits.}
\item
{In case of an adopted NFW dark halo the rotation curves
of the 8 most massive galaxies can be explained well, but in
some cases only for unrealistically small M/L ratios.}
\item
{For the four least luminous galaxies an NFW halo is clearly
inconsistent with the observations.}
\item
{The central regions of galaxies definitively have cores
and not cusps.}
\item
{When adopting a universal M/L ratio for all galaxies,
corrected for extinction and populations effects, the rotational
contribution of the luminous components becomes fixed a priori.}
\item
{For that assumption, in combination with a pseudo isothermal dark halo,
excellent fits to the observed rotation curves can be made.}
\item
{At the mean time, the rotational contribution of the disc is
then sub maximal being in agreement with
the observed stellar velocity dispersions.}
\item
{The core radii of the resultant halos are typically one to
a few kpc, apparently independent of galaxy mass.}
\item
{Such a universal M/L ratio generates very linear baryonic TF relations.}
\item
{In addition a ratio between dark and baryonic mass is then
determined which ranges between $\sim$ 9 for small galaxies
and $\sim$ 5 for large galaxies, at the outermost
measured rotation.}
\end{enumerate}


%

\appendix
\section{Tables}

\begin{table*}
\begin{minipage}{135mm}
\caption[]{Properties of whole galaxies}
\begin{tabular}{@{}lllllllll@{}}
\hline
       & Profile in & Opt. incl.     & \hi kin. & Dist. & $L_{\rm tot}$ & b/d   & Gas  &                           \\
Galaxy & band       & of disc ($^o$) & incl.    &       & observed      & ratio & mass & $M_{\rm gas}/L_{\rm tot}^R$ \\
       &            & ($q_0 = 0.11$) & ($^o$)   & (Mpc) & ($10^9 L^R_{\sun}$)     &       & ($10^9 M_{\sun}$) &    \\
\hline
N2841  & r          & 66             & 68       & 14.1  & 40.74         & 0.53  & 28.4     & 0.70  \\
N3992  & I          & 57             & 57       & 18.6  & 32.44         & 0.29  & 8.53     & 0.26  \\
N7331  & I          & ?              & 75       & 14.72 & 31.15         & 1.11  & 15.8    & 0.51  \\
N2998  & r          & 61             & 58       & 67.4  & 37.8          & 0     & 30.5  & 0.81  \\
N2903  & r          & 62             & 62       & 7.6   & 14.4          & 0     & 4.41     & 0.31  \\
N3198  & r          & 66             & 71       & 13.8  & 11.48         & 0     & 15.5     & 1.35  \\
N2403  & r          & 50             & 60       & 3.22  & 4.69          & 0     & 4.09     & 0.87  \\
N6503  & R          & 74             & 74       & 5.2   & 2.36         & 0     & 1.84     & 0.78  \\
N5585  & R          & 52             & 52       & 6.2   & 1.12         & 0     & 1.82     & 1.63  \\
N1560  & R          & 82             & 80       & 3.1   & 0.26         & 0     & 1.22    & 4.82  \\
N3109  & I          & 75             & 70       & 1.36  & 0.19          & 0     & 0.45     & 2.43  \\
D154   & R          & 56             & 64       & 3.8   & 0.029        & 0     & 0.43    & 15.1 \\
\hline
\end{tabular}
\end{minipage}
\end{table*}

\begin{table*}
\begin{minipage}{130mm}
\caption[]{Luminosities and extinction corrections in the R-band}
\begin{tabular}{@{}lllllllll@{}}
\hline
 & $L_{\rm tot}^{\rm obs}$ & Abs. Lum. & & & & & Abs. Lum. & $L_{\rm ec}^{\rm tot}$ \\
Galaxy & uncor. & uncor. & $\frac{a}{b}$ & $A^b$ & $A^{i-0}$ & $A^{i=0}$ & corrected & ext. corr. \\
 & $(10^9 L^R_{\sun})$ & (mag.) & & (mag.) & (mag.) & (mag.) & (mag.) & $(10^9 L^R_{\sun})$ \\
\hline
N2841b & 14.11 & -20.89 & 1    & 0.042 & 0.2$^{\ast}$ & 0    & -21.13 & 17.57 \\
N2841d & 26.63 & -21.58 & 2.58 & 0.042 & 0.61         & 0.26 & -22.49 & 61.60 \\
N3992b & 7.29  & -20.18 & 1    & 0.078 & 0.2$^{\ast}$ & 0    & -20.46 & 9.44  \\
N3992d & 25.15 & -21.52 & 1.79 & 0.078 & 0.358        & 0.25 & -22.21 & 47.25 \\
N7331b & 16.39 & -21.06 & 1    & 0.242 & 0.2$^{\ast}$ & 0    & -21.50 & 24.71 \\
N7331d & 14.76 & -20.94 & 3.57 & 0.242 & 0.783        & 0.25 & -22.22 & 47.64 \\
N2998  & 37.8  & -21.96 & 2.04 & 0.035 & 0.477        & 0.27 & -22.74 & 77.26 \\
N2903  & 14.4  & -20.92 & 2.08 & 0.083 & 0.409        & 0.23 & -21.64 & 28.05 \\
N3198  & 11.48 & -20.67 & 2.5  & 0.034 & 0.490        & 0.22 & -21.41 & 22.70 \\
N2403  & 4.69  & -19.70 & 1.54 & 0.107 & 0.177        & 0.17 & -20.15 & 7.11  \\
N6503  & 2.36  & -18.95 & 3.63 & 0.085 & 0.462        & 0.15 & -19.65 & 4.49  \\
N5585  & 1.12  & -18.14 & 1.61 & 0.042 & 0.111        & 0.10 & -18.39 & 1.41 \\
N1560  & 0.26  & -16.54 & 4.74 & 0.504 & 0.191        & 0.05 & -17.29 & 0.51 \\
N3109  & 0.19  & -16.19 & 3.57 & 0.178 & 0.047        & 0.02 & -16.44 & 0.23 \\
D154   & 0.029  & -14.16 & 1.75 & 0.025 & 0           & 0    & -14.19 & 0.029 \\
\hline
\multicolumn{9}{l}{\quad ${}^{\ast}$: estimated} \\
\end{tabular}
\end{minipage}
\end{table*}

\begin{table*}
\begin{minipage}{85mm}
\caption[]{Population corrections in the R-band}
\begin{tabular}{@{}lllll@{}}
\hline
 & & M/L & M/L & $L^{\rm tot}_{\rm epc}$ \\
Galaxy & $(B-V)^T_0$ & acc. to & scaled to &  \\
 & (RC3) & (B\&deJ) & $(B-V) = 0.6$ & $(10^9 L^R_{\sun})$ \\
\hline
N2814b & 1.05 & 4.20 & 3.55 & 62.30  \\
N2841d & 0.68 & 1.48 & 1.25 & 77.12  \\
N3992b & 0.94 & 3.08 & 2.60 & 24.56  \\
N3992d & 0.66 & 1.40 & 1.18 & 55.90  \\
N7331b & 1.0  & 3.65 & 3.08 & 76.13  \\
N7331d & 0.45 & 0.78 & 0.66 & 31.20  \\
N2998  & 0.48 & 0.84 & 0.71 & 55.09  \\
N2903  & 0.55 & 1.03 & 0.87 & 24.35  \\
N3198  & 0.43 & 0.73 & 0.62 & 14.07  \\
N2403  & 0.39 & 0.66 & 0.55 & 3.94   \\
N6503  & 0.57 & 1.09 & 0.92 & 4.13   \\
N5585  & 0.46 & 0.80 & 0.67 & 0.95   \\
N1560  & 0.57 & 1.09 & 0.92 & 0.47  \\
N3109  & 0.52 & 0.95 & 0.80 & 0.19  \\
D154   & 0.32 & 0.54 & 0.45 & 0.013 \\
\hline
\end{tabular}
\end{minipage}
\end{table*}

\begin{table*}
\begin{minipage}{105mm}
\caption[]{Maximum disc fits}
\begin{tabular}{@{}lllllll@{}}
\hline
Galaxy & $(M/L)_{\rm obs}$ & $(M/L)_{\rm epc}$ & $v^h_{\rm max}$ & Err. & $R^h_{\rm core}$ & Err. \\
 & $(M_{\sun}/L^R_{\sun})$ & $(M_{\sun}/L^R_{\sun})$ & (\kms) & (\kms) & (kpc) & (kpc) \\
\hline
N2841b & 5.67 & 1.28 & & & & \\
N2841d & 6.04 & 2.09 & 292.9 & 7.6  &  13.9 & 1.3  \\
N3992b & 6.48 & 1.92 &       &      &       &      \\
N3992d & 5.35 & 2.41 & 326.9 & 15.1 &  23.2 & 1.8  \\
N7331b & 4.31 & 0.93 &       &      &       &      \\
N7331d & 3.23 & 1.53 & 253.0 & 6.8  &  11.0 & 0.8 \\
N2998  & 4.36 & 2.99 & 254.9 & 55.4 & 32.5  & 10.9 \\
N2903  & 3.79 & 2.24 & 175.8 & 3.4  &  4.92 & 0.31 \\
N3198  & 3.96 & 3.23 & 153.1 & 2.9  & 11.23 & 0.64  \\
N2403  & 3.00 & 3.57 & 151.8 & 3.9  & 6.21  & 0.40  \\
N6503  & 4.37 & 2.50 & 124.7 & 2.1  & 4.33  & 0.22  \\
N5585  & 1.16 & 1.37 & 115.5 & 6.4  & 3.61  & 0.39  \\
N1560  & 5.44 & 2.96 & 135.8 & 9.4  & 7.95  & 0.73  \\
N3109  & 3.06 & 3.04 & 116.0 & 7.3  & 5.17  & 0.47  \\
D154   & 4.26 & 9.21 & 59.1   & 3.8   & 2.03  & 0.11  \\
\hline
\end{tabular}
\end{minipage}
\end{table*}

\begin{table*}
\begin{minipage}{110mm}
\caption[]{MOND fits}
\begin{tabular}{@{}llllll@{}}
\hline
Galaxy & $(M/L)_{\rm obs}$ & $(M/L)_{\rm epc}$ & Err. & $a_0$ & Err. \\
 & $(M_{\sun}/L^R_{\sun})$ & $(M_{\sun}/L^R_{\sun})$ & & (10$^{-8}$ cm s$^{-2}$) & (10$^{-8}$ cm s$^{-2}$) \\
\hline
N2841b &  5.67 &  1.28 & 0.13 &      &      \\
N2841d &  6.17 &  2.13 & 0.17 & 1.55 & 0.06 \\
N3992b &  7.20 &  2.14 & 0.21 &      &      \\
N3992d &  4.34 &  1.95 & 0.15 & 1.43 & 0.10 \\
N7331b &  4.49 &  0.97 & 0.07 &      &      \\
N7331d &  2.93 &  1.39 & 0.20 & 1.51 & 0.05 \\
N2998  &  4.08 &  2.80 & 0.16 & 0.53 & 0.06 \\
N2903  &  4.13 &  2.44 & 0.06 & 1.25 & 0.06 \\
N3198  &  3.87 &  3.16 & 0.06 & 0.53 & 0.01 \\
N2403  &  2.22 &  2.64 & 0.11 & 1.34 & 0.06 \\
N6503  &  3.82 &  2.19 & 0.05 & 1.09 & 0.03 \\
N5585  &  0.39 &  0.46 & 0.07 & 2.02 & 0.20 \\
N1560  &  1.30 &  0.71 & 0.07 & 1.01 & 0.03 \\
N3109  &  0.25 &  0.25 & 0.05 & 2.51 & 0.14 \\
D154   &  0.53 &  1.15 & 0.14 & 0.98 & 0.02 \\
\hline
\end{tabular}
\end{minipage}
\end{table*}

\begin{table*}
\begin{minipage}{125mm}
\caption[]{NFW-CDM$\Lambda$-AC fits}
\begin{tabular}{@{}lllllllll@{}}
\hline
Galaxy & $(M/L)_{\rm obs}$ & $(M/L)_{\rm epc}$ & Err. &
 $v^{\rm h, init}_{\rm max}$ & Err. & $R_s$ & c & $F_{200}$ \\
 & $(M_{\sun}/L^R_{\sun})$ & $(M_{\sun}/L^R_{\sun})$ & & (\kms) & (\kms) & (kpc) & & \\
\hline
N2841b & 3.52   & 0.80   & 0.10 &       &     &       &       &       \\
N2841d & 2.06   & 0.71   & 0.12 & 250.5 & 3.0 & 24.63 & 11.03 & 0.051 \\
N3992b & 2.25   & 0.67   & 0.09 &       &     &       &       &       \\
N3992d & 1.99   & 0.89   & 0.08 & 221.0 & 3.5 & 20.75 & 11.44 & 0.043 \\
N7331b & 2.54   & 0.55   & 0.02 &       &     &       &       &       \\
N7331d & 1.20   & 0.57 (fix) &   & 195.6 & 2.3 & 17.55 & 11.86 & 0.064 \\
N2998  & 1.28   & 0.88   & 0.12 & 178.4 & 6.1 & 15.47 & 12.18 & 0.091 \\
N2903  & 1.88   & 1.11   & 0.02 & 159.8 & 1.5 & 13.30 & 12.58 & 0.052 \\
N3198  & 0.85   & 0.70   & 0.04 & 139.0 & 1.0 & 10.99 & 13.11 & 0.065 \\
N2403  & 0.52   & 0.62   & 0.03 & 122.0 & 0.9 & 9.19  & 13.62 & 0.026 \\
N6503  & 1.46   & 0.84   & 0.03 & 104.0 & 0.7 & 7.39  & 14.27 & 0.035 \\
N5585  & $<$0   & $<$0   &      & 72.3  & 2.6 & 4.49  & 15.88 & 0.039 \\
N1560  & $<$0   & $<$0   &      & 58.9  & 1.0 & 3.39  & 16.86 & 0.050 \\
N3109  & $<$0   & $<$0   &      & 53.9  & 2.0 & 3.00  & 17.31 & 0.025 \\
D154   & $<$0   & $<$0   &      & 39.2  & 0.9 & 1.94  & 19.00 & 0.066 \\
\noalign{\smallskip}
N5585  & 0.85   & 1 (fix)&      & 52.7  &     & 2.91  & 17.42 & 0.163 \\
N1560  & 1.84   & 1 (fix)&      & 54.0  &     & 3.01  & 17.30 & 0.092 \\
N3109  & 1.01   & 1 (fix)&      & 50.8  &     & 2.77  & 17.61 & 0.042 \\
D154   & 0.46   & 1 (fix)&      & 38.8  &     & 1.91  & 19.06 & 0.071 \\
\hline
\end{tabular}
\end{minipage}
\end{table*}

\begin{table*}
\begin{minipage}{85mm}
\caption[]{Masses from observed stellar velocity dispersions
for $h/z_0 = 5$}
\begin{tabular}{@{}lllll@{}}
\hline
Galaxy & Mass & Err. & $(M/L)_{\rm epc}$ & Err. \\
 & $(10^9 M_{\sun})$ & $(10^9 M_{\sun})$ & $(M_{\sun}/L^R_{\sun})$ & $(M_{\sun}/L^R_{\sun})$ \\
\hline
N7331d & 17.7 & 7.6 & 0.57 & 0.24 \\
N3198  & 13.6 & 4.2 & 0.96 & 0.30 \\
N6503  & 3.3   & 1.2 & 0.81 & 0.29 \\
N2998  & 65.3 $\rightarrow$ 82.0 & 20\% & 1.19 $\rightarrow$ 1.49 & 0.27 \\
\hline
\end{tabular}
\end{minipage}
\end{table*}

\begin{table*}
\begin{minipage}{160mm}
\caption[]{$(M/L)_{\rm epc} = 1.0$, pseudo isothermal halo fits}
\begin{tabular}{@{}lllllllllll@{}}
\hline
  & \multicolumn{4}{l}{Direct fit} & \multicolumn{4}{l}{Fit to a-de-contracted halo} \\
Galaxy & $R_{\rm core}$ & Err. & $v_{\rm max}$ & Err. & $R_{\rm core}$ & Err. & $v_{\rm max}$ & Err. & ${\rho}_0^{\rm h}$ & Err. \\
 & (kpc) & (kpc) & (\kms) & (\kms) & (kpc) & (kpc) & (\kms) & (\kms) & ($M_{\odot}\; {\rm pc}^{-3}$) & ($M_{\odot}\; {\rm pc}^{-3}$) \\
\hline
N2841 & 1.90 & 0.27 & 256.7 & 2.3 & 6.39 & 0.52 & 267.6 & 3.7 & 0.032 & 0.005 \\
N3992 & 1.89 & 0.25 & 234.4 & 3.0 & 3.89 & 0.34 & 233.5 & 3.8 & 0.066 & 0.011 \\
N7331 & 7.02 & 0.57 & 230.1 & 4.9 & 19.46 & 1.79 & 268.6 & 14.2 & 0.0035 & 0.0007 \\
N2998 & 0.81 & 0.21 & 181.0 & 1.9 & 1.23 & 0.36 & 172.3 & 2.8 & 0.36 & 0.21 \\
N2903 & 0.54 & 0.13 & 174.5 & 1.8 & 1.07 & 0.17 & 171.1 & 2.2 & 0.47 & 0.15 \\
N3198 & 1.16 & 0.11 & 139.0 & 0.9 & 1.80 & 0.13 & 137.1 & 1.0 & 0.11 & 0.02 \\
N2403 & 1.24 & 0.06 & 130.1 & 1.0 & 1.64 & 0.08 & 130.2 & 1.2 & 0.12 & 0.01 \\
N6503 & 0.63 & 0.05 & 109.5 & 0.6 & 1.11 & 0.08 & 109.9 & 0.9 & 0.18 & 0.03 \\
N5585 & 2.69 & 0.26 & 107.0 & 4.9 & 2.87 & 0.30  & 106.3 & 3.7 & 0.025 & 0.006 \\
N1560 & 2.36 & 0.11 & 81.6  & 1.4 & 3.61 & 0.19 & 88.1  & 2.4 & 0.011 & 0.001 \\
N3109 & 2.85 & 0.15 & 88.5  & 2.2 & 3.82 & 0.09 & 100.4 & 1.4 & 0.013 & 0.001 \\
D154  & 1.18 & 0.05 & 51.8  & 2.7  & 1.44 & 0.08 & 53.2  & 3.4 & 0.025 & 0.002 \\
\hline
\end{tabular}
\end{minipage}
\end{table*}

\bsp
\label{lastpage}

\begin{thebibliography}{99}

\bibitem[]{}
Ahn, K., \& Shapiro, P.R. 2005, MNRAS, 363, 1092

\bibitem[Begeman(1987)]{b87}
Begeman, K. 1987, PhD Thesis, University of Groningen

\bibitem[]{}
Begeman, K. 1989, A\&A 223, 47

\bibitem[Begeman et al.(1991)]{bbs91}
Begeman, K.G., Broeils, A.H., \& Sanders, R.H. 1991, MNRAS, 249, 523 (BBS)

\bibitem[]{}
Bekenstein, J.D. 2004, Phys. Rev. D 70, 083509

\bibitem[Bell \& de Jong(2001)]{bdj01}
Bell, E. F., \& de Jong, R. S. 2001, ApJ 550, 212

\bibitem[Blais-Ouellette et al.(1999)]{bcac99}
Blais-Ouellette, S., Carignan, C., Amram, P., \& C\^{o}t\'e, S. 1999,  
AJ 118, 2123

\bibitem[Blais-Ouellette et al.(2001)]{bac01}
Blais-Ouellette, S., Amram, P., \& Carignan, C. 2001, AJ 121, 1952

\bibitem[]{}
Blais-Ouellette, S., Amram, P., Carignan, C., \& Swaters, R.
2004, A\&A, 420, 147

\bibitem[Blumenthal et al.(1986)]{bffp86}
Blumenthal, G. R., Faber, S. M., Flores, R., \& Primack, J. R. 1986,  
ApJ, 301, 27

\bibitem[Bond et al. 1996]{bond96}
Bond, J.R., Kofman, L., Pogosyan, D. 1996, Nature 380, 603

\bibitem[]{}
Bosma, A. 1978, PhD Thesis, University of Groningen

\bibitem[Bosma(1981)]{b81}
Bosma, A. 1981, AJ, 86, 1791

\bibitem[Bottema(1988)]{b88}
Bottema, R. 1988, A\&A, 197, 105

\bibitem[Bottema(1989)]{b89}
Bottema, R. 1989, A\&A, 221, 236

\bibitem[Bottema(1993)]{b93}
Bottema, R. 1993, A\&A, 275, 16

\bibitem[]{}
Bottema, R. 1997, A\&A, 328, 517

\bibitem[Bottema(1999)]{b99}
Bottema, R. 1999, A\&A, 348, 77

\bibitem[Bottema(2002)]{b02}
Bottema, R. 2002, A\&A, 388, 809

\bibitem[]{}
Bottema, R. 2003, MNRAS, 344, 358

\bibitem[]{}
Bottema, R., \& Kregel, M. 2014, in preparation

\bibitem[Bottema \& Verheijen(2002)]{bv02}
Bottema, R., \& Verheijen, M. A. W. 2002, A\&A 388, 793

\bibitem[Bottema et al.(2002)]{bprs02}
Bottema, R., Pesta\~{n}a, J.L.G., Rothberg, B., \& Sanders, R.H.  
2002, A\&A 393, 453 (BPRS)

\bibitem[]{}
Broeils, A.H. 1992a, PhD. Thesis, University of Groningen

\bibitem[Broeils(1992)]{b92}
Broeils, A.H. 1992b, A\&A, 256, 19

\bibitem[]{}
Bullock, J.S., Kolatt, T.S., Sigad, Y., et al. 2001, MNRAS, 321, 559

\bibitem[]{}
Burkert, A. 1995, ApJ, 447, L25

\bibitem[Carignan \& Beaulieu(1989)]{cb89}
Carignan, C., \& Beaulieu, S. 1989, ApJ 347, 760

\bibitem[]{}
Carignan, C., \& Freeman, K.C. 1988, ApJ, 332, L33

\bibitem[Carignan \& Purton(1998)]{cp98}
Carignan, C., \& Purton, C. 1998, ApJ 506, 125 (CP98)

\bibitem[]{}
Casertano, S. 1983, MNRAS, 203, 735

\bibitem[]{}
Casertano, S., \& van Gorkom, J.H. 1991, AJ 101, 1231

\bibitem[Choi et al.(2006)]{clmw06}
Choi, J.-H., Lu, Y., Mo, H. J., \& Weinberg, M. D. 2006, MNRAS, 372,  
1869

\bibitem[]{}
Col\'{\i}n, P., Avila-Reese, V., Valenzuela, O., \& Firmani, C.
2002, ApJ, 581, 777

\bibitem[C\^{o}t\'e et al.(1991)]{ccs91}
C\^{o}t\'e, S., Carignan, C., \& Sancisi, R. 1991, AJ 102, 904

\bibitem[]{}
Courteau, S., \& Rix, H.W. 1999, ApJ 513, 561

\bibitem[]{}
Dav\'e, R., Spergel, D.N., Steinhardt, P.J., \& Wandelt, B.D. 2001,
ApJ, 547, 574

\bibitem[Davis et al. 1985]{davis85}
Davis, M., Efstathiou, G., Frenk, C.S., White, S.D.M.
1985, ApJ 292, 371

\bibitem[de Blok \& Bosma(2002)]{dbb02}
de Blok, W. J. G., \& Bosma, A. 2002, A\&A 385, 816

\bibitem[]{}
de Blok, W.J.G., \& McGaugh, S.S. 1998, ApJ 508, 132

\bibitem[de Blok et al.(2001a)]{dbmbr01}
de Blok, W.J.G., McGaugh, S.S., Bosma, A., \& Rubin, V.C. 2001a,  
ApJ, 552, L23

\bibitem[de Blok et al. (2001b)]{dbmr01}
de Blok, W.J.G., McGaugh, S.S., \& Rubin, V.C. 2001b,
AJ, 122, 2396

\bibitem[]{}
de Blok, W.J.G., Bosma, A., \& McGaugh, S.S. 2003, MNRAS 340, 657

\bibitem[]{}
de Blok, W.J.G., Walter, F., Brinks, E., et al. 2008, AJ, 136, 2648

\bibitem[]{}
de Jong, R.S. 1996, A\&AS, 118, 557

\bibitem[]{}
de Vaucouleurs, G., de Vaucouleurs, A., Corwin, H.G., et al. 1991,
Third Reference Catalogue of Bright Galaxies (Springer verlag, New York),
(RC3)

\bibitem[]{}
Diemand, J., Moore, B., \& Stadel, J. 2004, MNRAS, 353, 624

\bibitem[Drozdovsky \& Karachentsev(2000)]{dk00}
Drozdovsky, I. O., \& Karachentsev, I. D. 2000, A\&AS 142, 425

\bibitem[]{}
Dutton, A.A., Courteau, S., de Jong, R.S., \& Carignan, C. 2005, ApJ 619, 218

\bibitem[]{}
Firmani, C., D'Onghia, E., Chincarini, G., Hern\'andez, X., \&
Avila-Reese, V. 2001, MNRAS, 321, 713

\bibitem[Freedman \& Madore(1988)]{fm88}
Freedman, W. L., \& Madore, B. F. 1988, ApJ 332, L63

\bibitem[Freedman et al.(2001)]{fmg+01}
Freedman, W. L., Madore, B. F., Gibson, B. K., et al. 2001, ApJ, 553, 47

\bibitem[]{}
Garrison-Kimmel, S., Rocha, M., Boylan-Kolchin, M. \& al. 2013, MNRAS,
433, 3539

\bibitem[]{}
Gentile, G., Famaey B. \& de Blok, W.J.G. 2010, A\&A, 527, A76+

\bibitem[]{}
Gentile, G., Salucci, P., Klein, U., Vergani, D., \& Kalberla, P.
2004, MNRAS 351, 903

\bibitem[Gnedin et al.(2004)]{gkkn04}
Gnedin, O. Y., Kravtsov, A. V., Klypin, A. A., \& Nagai, D. 2004, ApJ,  
616, 16

\bibitem[]{}
Governato, F., Brook, C., Mayer, L., et al. 2010, Nature, 463, 203

\bibitem[]{}
Herrmann, K.A. \& Ciardullo, R. 2009, ApJ, 705, 1686

\bibitem[Hernquist(1990)]{h90}
Hernquist, L. 1990, ApJ, 356, 359

\bibitem[]{}
Hoekstra, H., van Albada, T.S., \& Sancisi, R. 2001, MNRAS, 323, 453

\bibitem[Hughes et al.(1998)]{hhh+98}
Hughes, S. M. G., Han, M., Hoessel, J., et al. 1998, ApJ, 501, 32

\bibitem[Jesseit et al.(2002)]{jnb02}
Jesseit, R., Naab, T., \& Burkert, A. 2002, ApJ, 571, L89

\bibitem[]{}
Jing, Y.P. 2000, ApJ, 535, 30

\bibitem[Jobin \& Carignan(1990)]{jc90}
Jobin, M., \& Carignan, C. 1990, AJ 100, 648

\bibitem[Karachentsev \& Sharina(1997)]{ks97}
Karachentsev, I. D., \& Sharina, M. E. 1997, A\&A 324, 457

\bibitem[Kelson et al.(1999)]{kis+99}
Kelson, D. D., Illingworth, G. D., Saha,  A., et al. 1999, ApJ, 514, 614

\bibitem[]{}
Kennicutt, R.C. Jr., Armus, L., Bendo, G, et al. 2003, PASP, 115, 928

\bibitem[]{}
Kent, S.M. 1986, AJ, 91, 1301 (K86)

\bibitem[Kent(1987)]{k87}
Kent, S.M. 1987, AJ, 93, 816 (K87)

\bibitem[]{}
Knapp, G., \& Kormendy, J. (eds) 1987, in IAU Symp. No. 117, Dark matter
in the universe (Reidel Pub. Dordrecht) 

\bibitem[Koopmans et al. (2006)]{koop+06}
Koopmans, L.V.E., Treu, T., Bolton, A.S., Burles, S., Moustakas, L.A. 2006,
ApJ 649, 599

\bibitem[]{}
Kregel, M., van der Kruit, P.C., \& Freeman, K.C. 2005, MNRAS 358, 503

\bibitem[Krismer et al.(1995)]{ktg95}
Krismer, M., Tully, R. B., Gioia, I. M. 1995, AJ 110, 1584

\bibitem[Kroupa (2001)]{kroupa01}
Kroupa, P. 2001, MNRAS, 322, 231

\bibitem[Lake(1989)]{l89}
Lake, G. 1989, ApJ, 345, L17

\bibitem[Lauberts \& Valentijn(1989)]{lv89}
Lauberts, A., \& Valentijn, E. A. 1989, The Surface Photometry  
Catalogue of the ESO-
Uppsala Galaxies, (European Southern Observatory, Garching), p. 10

\bibitem[Lee \& Madore(1993)]{lm93}
Lee, M. G., \& Madore, B. F. 1993, AJ 106, 964

\bibitem[Longo \& de Vaucouleurs(1983)]{ldv83}
Longo, G., \& de Vaucouleurs, A. 1994, A General Catalogue
of Photoelectric Magnitudes and Colors in the U, B, V system of
3578 Galaxies Brighter than the 16th Magnitude, University of
Texas Monographs in Astronomy, No. 3

\bibitem[]{}
Ludlow, A.D., Navarro, J.F., Angulo, R.E., et al. 2014, MNRAS, 441, 378

\bibitem[]{}
Lynden-Bell, D., \& Eggleton, P.P. 1980, MNRAS, 191, 483

\bibitem[]{}
Macci\`o, A.V., Dutton, A.A., van den Bosch, F.C., et al.
2007, MNRAS, 378, 55

\bibitem[]{}
Macci\`o, A.V., Dutton, A.A. \& van den Bosch, F.C. 2008, MNRAS, 391, 1940

\bibitem[Macri et al.(2001)]{msbfgjmr01}
Macri, L. M., Stetson, P. B., Bothun, G. D., Freedman, W. L.,  
Garnavich, P. M., Jha, S. Madore, B. F., \& Richmond, M. W. 2001, ApJ,  
559, 243

\bibitem[Marcelin et al.(1983)]{mbg83}
Marcelin, M., Boulesteix, J., \& Georgelin, Y. 1983, A\&A 128, 140

\bibitem[]{}
Martinsson, T.P.K., Verheijen, M.A.W., Westfall, K.B., et al. 2013,
A\&A, 557, A131

\bibitem[McGaugh \& de Blok(1998)]{mdb98}
McGaugh, S. S., \& de Blok, W. J. G. 1998, ApJ, 499, 66 

\bibitem[McGaugh et al.(2000)]{msvdb00}
McGaugh, S.S., Schombert, J.M., Bothun, G.D., \& de Blok, W.J.G.  
2000, ApJ, 533, L99

\bibitem[Milgrom(1983)]{m83}
Milgrom, M. 1983 ApJ, 270, 365

\bibitem[Milgrom(1988)]{m88}
Milgrom, M. 1988, ApJ, 333, 689

\bibitem[]{}
Moore, B., Quinn, T., Governato, F., Stadel, J., \& Lake, G. 1999,
MNRAS, 310, 1147

\bibitem[Musella et al.(1997)]{mpg97}
Musella, I., Piotto, G., \& Capaccioli, M. 1997, AJ 114, 976

\bibitem[]{}
Navarro, J.F. 1998, [astro-ph/9807084]

\bibitem[Navarro et al.(1997)]{nfw97}
Navarro, J.F., Frenk, C.S., \& White, S.D.M. 1997, ApJ, 490, 493 (NFW)

\bibitem[]{}
Neto, A.F., Liang, G., Bett, P.B., et al. 2007, MNRAS, 381, 1450

\bibitem[]{}
Noordermeer, E. 2006, PhD. Thesis, University of Groningen

\bibitem[Noordermeer \& Verheijen(2007)]{nv07}
Noordermeer, E., \& Verheijen, M. A. W. 2007, MNRAS, 381, 1463

\bibitem[]{}
Pe\~narrubia, J., Pontzen, A., Walker, M.G. \& al. 2012, ApJ, 759, L42

\bibitem[]{}
Persic, M., Salucci, P. \& Stel, F. 1996, MNRAS, 281, 27

\bibitem[]{}
Pontzen, A. \& Governato, F. 2012, MNRAS, 421, 3464

\bibitem[Prada et al.(1996)]{pgpm96}
Prada, F., Gutierrez, C. M., Peletier, R. F., \& McKeith, C. D. 1996,  
ApJ, 463, L9

\bibitem[]{}
Randriamampandry, T. \& Carignan, C. 2014, MNRAS, 439, 2132

\bibitem[]{}
Read, J.I. \& Gilmore, G. 2005, MNRAS, 356, 107

\bibitem[Reed(1989)]{r89}
Reed, B.C. 1989, Am. J. Phys., 57 (7), 642

\bibitem[Rubin et al.(1965)]{rbbcp65}
Rubin, V. C., Burbidge, E. M., Burbidge, G. R., Crampin, D. J., \&  
Prendergast, K. H. 1965, ApJ, 141, 759

\bibitem[Rubin et al.(1985)]{rbft85}
Rubin, V. C., Burstein, D., Ford, W. K., Jr., \& Thonnard, N. 1985,  
ApJ, 289, 81

\bibitem[]{}
Salpeter, E.E. 1955, ApJ 121, 161

\bibitem[]{}
Salucci, P., Walter, F., \& Borriello, A. 2003, A\&A, 409, 53

\bibitem[]{}
Sancisi, R., \& van Albada, T.S. 1987, in IAU Symp. 117, Dark matter
in the universe, ed. J. Kormendy, \& G. Knapp (Reidel Pub. Dordrecht),
67

\bibitem[]{}
Sanders, R.H. 1990, A\&A Rev., 2, 1

\bibitem[Sanders(1996)]{s96}
Sanders, R. H. 1996, ApJ, 473, 117

\bibitem[]{}
Sanders, R. H., \& McGaugh S. S. 2002, ARA\&A, 40, 263

\bibitem[Sanders \& Verheijen(1998)]{sv98}
Sanders, R. H., \& Verheijen, M. A. W. 1998, ApJ, 503, 97

\bibitem[Schlegel et al.(1998)]{sfd98}
Schlegel, D. J., Finkbeiner, D. P., \& Davis, M. 1998, ApJ 500, 525

\bibitem[]{}
Silk, J. 1987, in A Unified View of the Macro- and Micro-Cosmos,
eds. A. de Rujula, D.V. Nanopoulos, P.A. Shaver (World Scientific,
Singapore), p. 277

\bibitem[]{}
Simon, J.D., Bolatto, A.D., Leroy, A., \& Blitz, L. 2003,
ApJ 596, 957

\bibitem[Soszy\'nski et al.(2006)]{sgpbks06}
Soszy\'nski, I., Gieren, W., Pietrzy\'nski, G., Bresolin, F.,  
Kudritzki, R.-P., \& Storm, J. 2006, ApJ, 648, 375

\bibitem[]{}
Spergel D.N., Bean, R., Dor\'e, M.R., et al. 2007, ApJS, 170, 377

\bibitem[]{}
Spergel, D.N., \& Steinhardt, P.J. 2000, Phys. Rev. Lett., 84, 3760

\bibitem[]{}
Swaters, R.A. 1999, PhD. Thesis, University of Groningen

\bibitem[Swaters \& Balcells(2002)]{sb02}
Swaters, R.A., \& Balcells, M. 2002, A\&A 390, 863

\bibitem[]{}
Swaters, R.A., Sanders, R.H. \& McGaugh, S.S. 2010, ApJ, 718, 380

\bibitem[Thuan \& Gunn(1976)]{tg76}
Thuan, T.X., \& Gunn, J.E. 1976, PASP 88, 543

\bibitem[]{}
Trimble, V. 1987, ARA\&A 25, 425

\bibitem[Tully \& Fisher(1977)]{tf77}
Tully, R. B., \& Fisher, J. R. 1977, A\&A 54, 661

\bibitem[Tully \& Fouqu\'e(1985)]{tf85}
Tully, R. B., \& Fouqu\'e, P. 1985, ApJS 58, 67

\bibitem[Tully \& Pierce(2000)]{tp00}
Tully, R. B., \& Pierce, M. J. 2000, ApJ 533, 744

\bibitem[Tully et al.(1996)]{tvphw96}
Tully, R. B., Verheijen, M. A. W., Pierce, M. J., Huang, J.-S., \&  
Wainscoat, R. J. 1996, AJ, 112, 2471

\bibitem[Tully et al.(1998)]{tphsvw98}
Tully, R. B., Pierce, M. J., Huang, J.-S., Saunders, W., Verheijen, M.  
A. W., \& Witchalls, P. L. 1998, AJ, 115, 2264

\bibitem[]{}
van Albada, T.S., \& Sancisi, R. 1986, Phil. Trans. R. Soc. London,
Ser. A 320, 447

\bibitem[van Albada et al.(1985)]{vabbs85}
van Albada, T. S., Bahcall, J. N., Begeman, K., \& Sancisi, R. 1985,  
ApJ, 295, 305

\bibitem[]{}
van der Kruit, P.C., \& Searle, L. 1981, A\&A, 95, 105

\bibitem[]{}
van der Kruit, P.C., \& Searle, L. 1982, A\&A, 110, 61
\bibitem[Verheijen(2001)]{v01}
Verheijen, M. A. W. 2001, ApJ, 563, 694
\bibitem[van Albada et al.(1986)]{wvdka86}
Wevers, B. M. H. R., van der Kruit, P. C., \& Allen, R. J. 1986, A 
\&AS, 66, 505
\bibitem[]{}
Yoshida, N., Springel, V., White, S.D.M., \& Tormen, G. 2000,
ApJ, 535, L103

\end{thebibliography}
\end{document}